\documentclass[pra,twocolumn,superscriptaddress,showpacs,amsmath,amssymb]{revtex4-1}

\pacs{42.50.Nn, 03.65.Yz, 05.70.Ln, 31.15.xk}

\usepackage{latexsym}
\usepackage{amsmath, amsthm, amssymb} 
\usepackage{txfonts}

\usepackage{comment}

\usepackage[latin1]{inputenc}
\usepackage{amsmath}
\usepackage{amsfonts}
\usepackage{amssymb}
\usepackage{graphics}
\usepackage{setspace}
\usepackage[usenames]{color}
\usepackage{array}

\def \widetilde#1{{#1}}

\def \MARKOV {{}}
\def \EQUILIB {{EQ}}

\usepackage[latin1]{inputenc}
\usepackage{amsmath}
\usepackage{amsfonts}
\usepackage{amssymb}
\usepackage{graphicx}
\usepackage{setspace}
\usepackage[usenames]{color}
\usepackage{epstopdf}
\usepackage{array}

\newcommand{\be}{\begin{equation}}
\newcommand{\ee}{\end{equation}}

\def \be{\begin{equation}}
\def \ee{\end{equation}}
\def \ba{\begin{array}}
\def \ea{\end{array}}
\def \bea{\begin{eqnarray}}
\def \eea{\end{eqnarray}}
\def \nn{\nonumber}
\def \half{{1\over 2}}

\def \e{{\epsilon}}

\def \a{{\alpha}}

\def \b{{\beta}}

\def \w{{\omega}}

\def \x{{\chi}}
\def \e{{\epsilon}}

\def \G{{\Gamma}}

\def \yd{^\dagger}
\def \av#1{{\langle#1\rangle}}

\def \beas{\begin{eqnarray*}}
\def \eeas{\end{eqnarray*}}

\def \half{{\frac{1}{2}}}

\newcounter{indice}

\def \bn{\begin{enumerate}}
\def \en{\end{enumerate}}

\def \bb{\begin{thebibliography}{99}}
\def \eb{\end{thebibliography}}

\begin{document}
\title{Keldysh approach for non-equilibrium phase transitions in quantum optics:\\ beyond the Dicke model in optical cavities}

\author{Emanuele G. Dalla Torre}
\email{emanuele@physics.harvard.edu}
\affiliation{Department of Physics, Harvard University, Cambridge MA 02138}

\author{Sebastian Diehl}
\affiliation{Institute for Quantum Optics and Quantum Information of the Austrian Academy of Sciences, A-6020 Innsbruck, Austria}
\affiliation{Institute for Theoretical Physics, University of Innsbruck, A-6020 Innsbruck, Austria}

\author{Mikhail D. Lukin}
\affiliation{Department of Physics, Harvard University, Cambridge MA 02138}

\author{Subir Sachdev}
\affiliation{Department of Physics, Harvard University, Cambridge MA 02138}

\author{Philipp Strack}
\affiliation{Department of Physics, Harvard University, Cambridge MA 02138}

\date{\today}

\begin{abstract}
We investigate non-equilibrium phase transitions for driven atomic ensembles, interacting with a cavity mode, coupled to a Markovian dissipative bath. In the thermodynamic limit and at low-frequencies, we show that the distribution function of the photonic mode is thermal, with an effective temperature set by the atom-photon interaction strength. This behavior characterizes the static and dynamic critical exponents of the associated superradiance transition. Motivated by these considerations, we develop a general Keldysh path integral approach, that allows us to study physically relevant nonlinearities beyond the idealized Dicke model. Using standard diagrammatic techniques, we take into account the leading-order corrections due to the finite number of atoms $N$. For finite $N$, the photon mode behaves as a damped, classical non-linear oscillator at finite temperature. For the atoms, we propose a Dicke action that can be solved for any $N$ and correctly captures the atoms' depolarization due to dissipative dephasing.
\end{abstract}

\maketitle

\section{Introduction}
\label{sec:intro}

Much interest has recently been directed towards understanding many-body dynamics  in open systems away from thermal equilibrium. 
This subject is not new, as the analogies between threshold phenomena in 
dynamical systems, such as the laser, and the conventional phase transitions have been recognized over 40 years ago. However,   
recent experiments with ultracold atoms in optical cavities offer intriguing possibilities to explore the physics of strongly interacting 
atom-photon systems far away from thermal equilibrium from a new vantage point. Many fundamental concepts of condensed matter 
physics, ranging from classification of phase transitions to the universal behavior of correlation functions in the vicinity of quantum critical points in the presence of driving and dissipation, need to be revisited in light of these developments. 

In this paper, we investigate non-equilibrium phase transitions for driven atomic ensembles interacting with a cavity mode that is subject to dissipation, focusing specifically on 
the dynamical superradiance transitions and associated self-organization of the atoms observed in Refs.~\cite{black03,baumann10,bohnet12}. Due to the interplay of external driving, Hamiltonian dynamics and dissipative 
processes, the observed Dicke superradiance transitions \cite{black03,baumann10} exhibit several properties \cite{dimer07,nagy11,oeztop11,bhaseen12} which are not present in the closed Dicke model \cite{hepp73,wang73,popov82,emary03}. Recently, it was shown 
that other more interesting quantum many-body phases, such as quantum spin and charge glasses with 
long-range, random interactions \cite{strack11,gopa11,muller12} mediated by multiple photon modes, could potentially 
be simulated with many-body cavity QED. 

 
In what follows we first review the physics of the non-equilibrium Dicke transition in optical cavities with conventional techniques of quantum optics. Using linearized Heisenberg-Langevin equations, we demonstrate that at low frequencies, close to the phase transition, 
the system evolves into a thermal state with a high effective temperature, proportional to the atom-photon interaction strength. 
The static and dynamic critical exponents at the high-temperature phase transitions are analogous to those in a 
conventional laser (or, more precisely, optical parametric oscillator) threshold. 


To treat the interplay between external driving, dissipation and many-body interactions in a more general setting, we next develop a 
unified approach to describe phase transitions in open quantum systems based on  
 Keldysh path integrals \cite{castroneto90,szymanska06,szymanska07,sarang10,kamenev09,kamenev_book,altlandbook}. This approach is used to analyze the driven Dicke model in the presence of finite-size effects and atomic dissipative processes. Both perturbations are non-linear and cannot be treated by the usual quantum-optical methods. Instead, we apply non-perturbative techniques, specific to the path-integral approach. We find that the low-frequency dynamics is thermal even in this case, allowing for an effective equilibrium description.

We expect the Keldysh approach
to be directly applicable to other dissipative models such as the recently discussed central spin model \cite{kessler12} or fermionic lattice models \cite{PhysRevLett.105.227001,eisert10,hoening12,horstmann12}. 
We believe the Keldysh calculations are not more involved, and sometimes simpler, than those of the usual quantum optics frameworks \cite{zollerbook, scully97}. At the same time, they 
facilitate an easy comparison to other phase transitions of condensed matter physics.
One of the objectives of the present 
paper is to make the Keldysh approach more accessible to the broader quantum optics community. At the same time, we hope that the 
Keldysh perspective will be helpful for condensed matter physicists to understand driven dissipative atom-photon systems--especially in view of the qualitatively different energy scales and bath properties in quantum optics.

The paper is organized as follows. In Sec.~\ref{sec:review} we introduce the Dicke model and perform a brief analysis with linearized Heisenberg-Langevin equations pointing out that the relevant low-frequency correlations are thermal. In Sec.~\ref{sec:cavityonly}, we map the operators of the master equation and the associated Liouvillians for dissipative processes to the field content of an equivalent real-time, dissipative Keldysh action $S[a^*,a]$ with $a^*$, $a$ the photon field variables. In Sec.~\ref{sec:dicke_kappa} we introduce the atomic degrees of freedom and study the thermodynamic limit $N\to\infty$ of the open Dicke model. We define a non-equilibrium distribution function $F(\omega)$ and compare it to the equilibrium case, finding that both diverge at low-frequencies as $1/\omega$. In Sec.~\ref{sec:finite_N} we study the effects of a finite size $N$. Combining analytic and numerical methods, we derive the critical scaling of the photon number as function of $N$, and find it to be equivalent to an equilibrium system at finite temperature and distinct from the zero temperature case. In Sec.~\ref{sec:spontem} we propose an effective method to
describe
the effects of single atom decay across the phase transition of the Dicke model. 
We again find that the distribution function is thermal, but with renormalized couplings and, in particular, a different critical coupling $g_c$. Sec. \ref{sec:conclusion} concludes the paper with a summary of our main results and some final remarks.

\section{Thermal nature of the open Dicke transition}
\label{sec:review}


The Dicke Hamiltonian \cite{wang73,hepp73} describes $N$ two-level systems or ``qubits'' represented by Pauli matrix operators
$\sigma^{x}_{i}$, $\sigma^{z}_i$ coupled to a quantized photon mode represented by bosonic creation and annihilation 
operators $\hat{a}^\dagger$, $\hat{a}$:
\be 
H = \w_0 \hat a\yd \hat a +\frac{\w_z}2 \sum_{i=1}^N \sigma^z_i + \frac{g}{\sqrt{N}} \sum_{i=1}^N \sigma_i^x \left(\hat a\yd + \hat{a}\right)\;.\label{eq:HDicke1}
\ee
Here $\omega_0$ is the photon frequency, $\omega_z$ the level-splitting of the qubits, and $g$ the qubit-photon coupling,
assumed to couple all qubits uniformly to the photon. 
Eq.~(\ref{eq:HDicke1}) is invariant under an Ising-type $Z_2$-transformation, $\hat{a}\rightarrow -\hat{a}$ and $\sigma^x_i\rightarrow -\sigma^x_i$. In the thermodynamic limit $N\rightarrow \infty$, and for sufficiently strong qubit-photon coupling $g$, the ground state of Eq.~(\ref{eq:HDicke1}) spontaneously breaks this Ising symmetry and exhibits a phase transition to a ``superradiant'' phase with a photon condensate $\langle \hat a \rangle$. 

In the context of ultracold dilute gases in optical cavities, 
Dimer {\it et al.} \cite{dimer07} proposed to implement the qubits using two hyperfine states of the atoms and showed that, close to the transition, the relevant Hilbert space can be exactly mapped to Eq.~(\ref{eq:HDicke1}). Inspired by the work of Dimer {\it et al.} \cite{dimer07}, 
the qubit states of the Dicke model were realized using two collective motional degrees of the Bose-Einstein condensate of the atoms in the cavity \cite{nagy10,baumann10}, see \cite{ritschratsch12} for a review. In that case $\omega_z$ becomes a collective recoil frequency and the two Dicke states are components of a dynamically forming charge density wave. 

This open realization of the Dicke model in optical cavities with pumped atoms is different from the closed system Dicke model Eq.~(\ref{eq:HDicke1}), due to the interplay of coherent drive and dissipation:

{\em 1. Coherent drive:} 
The photon-atom coupling $g$ describes the scattering of pump-photons and rotates, as function of time, at the pump frequency $\w_p$. To obtain the time-independent Dicke model (\ref{eq:HDicke1}), one has to move to a rotating frame, where the explicit time-dependence of the original Hamiltonian  is ``gauged'' away\footnote{See for example Eq.~(4) of Ref.~\cite{dimer07} or Eq.~(2) of Ref.~\cite{maschler08}}. In this frame, the parameter $\w_0$ appearing in Eq.~(\ref{eq:HDicke1}) is the cavity-pump detuning $\omega_0=\omega_c - \omega_p$, where $\omega_c$ the bare cavity frequency. 

{\em 2. Dissipation:}
In addition to the coherent dynamics generated by the Hamiltonian Eq.~(\ref{eq:HDicke1}), there is a dissipative contribution consisting of cavity loss and dissipative processes for the atoms. In the rotating frame, this vacuum is effectively out-of-equilbrium, and leads to the non-equilibrium Markovian master equation (cf. Appendix \ref{app:integration}), 
\begin{align}
\partial_t \rho &=  -\mathrm i [H, \rho] + \mathcal{L}~\rho\;,
\label{eq:master-gen} 
\end{align}
Here $\rho$ is the density matrix and ${\mathcal L}$ the Liouville operator in Lindblad form
\be
\mathcal{L}~\rho = \sum_\alpha \kappa_\alpha \Big(2 L_\alpha \rho L_\alpha^\dag -  \{L_\alpha^\dag L_\alpha , \rho\} \Big)\;\label{eq:master-gen2},
\ee
where the curly brackets  $\{,\}$ denotes the anti-commutator and $L_\alpha$ is a set of Lindblad or quantum jump operators. In the present work we consider two types of disspative processes: cavity photon loss and single atom dissipative dephasing. The former is modeled by the Liouvillian: 
\be
\mathcal{L}_\text{cav}~\rho = \kappa(2 \hat a \rho \hat a^\dag - \{\hat a^\dag \hat a , \rho\} )\;,
\label{eq:cav_liou}
\ee
where $\kappa$ is an effective decay rate (inverse lifetime) of a cavity photon of the order a few MHz \cite{baumann10}.
Modelling the dissipative dynamics of the atoms depends on the specific implementation of the driven Dicke model usually involving 
local processes of each two-level atom separately. In Sect. \ref{sec:spontem}, we account for dissipative dephasing of the atoms 
in an approximate way by resorting to a simplified effective low frequency model.

%
%
%

\subsection{Heisenberg-Langevin analysis}

\label{sec:prelim}

\def\mF{{\mathcal{F}}}

We now study the above non-equilibrium Dicke model using conventional quantum optical techniques, namely,  the Heisenberg-Langevin equations of motion.  We will later repeat and extend these calculations using the Keldysh path integral approach, in the following Sections. The master equation (\ref{eq:master-gen}--\ref{eq:master-gen2}) with Hamiltonian (\ref{eq:HDicke1}) and cavity dissipation (\ref{eq:cav_liou}) is equivalent to the equations of motion:
\bea
\dot{\hat{a}}  &=& -i\w_0 \hat{a} - \kappa \hat{a} - \frac{i g}{\sqrt{N}} \sigma^x_i + \mF,\nn\\
\dot \sigma^+_i  &=& i\w_z \sigma^{+}_i - \frac{i g}{\sqrt{N}} \sigma^z_i (\hat{a}+\hat{a}\yd),\nn\\
\dot \sigma^z_i &=& -2\frac{i g}{\sqrt{N}} (\sigma^+_i - \sigma^-_i)(\hat{a}+\hat{a}\yd).\label{eq:HL}
\eea
Here the force $\mF=\mF(t)$ is a stochastic Markovian operator satisfying $\av{\mF(t)\mF\yd(t')} = 2\kappa\delta(t-t')$ and $\av{\mF(t)\yd\mF(t)}=0$. This term is needed in order to preserve the commutation relation $[\hat{a}(t),\hat{a}\yd(t)]=1$, which would otherwise exponentially decrease. 
See for example Ref.~\cite{sachdev84} for a detailed study of the single-atom case, $N=1$.
%

To analyze the dynamics below the superradiance threshold in the limit of $N\to\infty$, we assume that the atoms are fully polarized 
$S^z=\half\sum_i \sigma^z_i \approx - N/2$, and neglect non-linear terms in the equations of motion. The resulting 
operator equations can be solved exactly in the Fourier domain as has been done in detail in the work of Dimer {\em et al.} \cite{dimer07} and 
we will not repeat their calculations here. We just add one simple point to their comprehensive analysis, namely that 
the relevant low-frequency dynamics of the photons occurs in the presence of a finite effective temperature. 
This is most easily seen 
from the equation of motion for the real-valued phase-space coordinate $x(\omega) =(\hat{a}(\omega)+\hat{a}\yd(-\omega))/\sqrt{2\w_0}$.
Defining stochastic force operators $f(\w) = \frac{1}{\sqrt{2\w_0}}\left[\mF(\w)\left(\kappa-i(\w_0+\w)\right) + \mF\yd(-\w) \left(\kappa+i(\w_0-\w)\right)\right]$, we can write the 
%
equation of motion as
\be
\left((-\kappa+i\w)^2 + \w_0^2 - \frac{4g^2\w_z\w_0}{\w_z^2-\w^2}\right) x(\omega) = f(\w)\;.\label{eq:langevin_misha} 
\ee
The force operator satisfies
\be
 \tfrac{1}{2}\av{f(\w)f(\w') + f(\w')f(\w)}= \kappa\frac{\kappa^2+\w_0^2+\w^2}{\w_0}\delta(\w+\w')\;.
 \label{eq:ff}
\ee

At low frequencies,we can neglect high-order terms in $\w$. Eq.~(\ref{eq:langevin_misha}) becomes identical to the Langevin equation of a classical particle in a harmonic potential with oscillation frequency $\a$, defined by
\begin{align}
\a^2 &= \kappa^2 + \w_0^2- \frac{4g^2\w_0}{\w_z} \;,
\label{eq:alpha}
\end{align}
and friction constant $2\kappa$.
In the same low-frequency approximation, the correlation function of the stochastic force operators on the right-hand-side of Eq.~(\ref{eq:ff}) becomes identical to the ``noise correlations"
provided by an equilibrium classical bath at a non-zero temperature
\bea
\widetilde{T^{\rm eff}_{x}}&=& \frac{\w_0^2+\kappa^2}{4\w_0}=\frac{g_c^2}{\omega_z}\;.
\label{eq:Teff2}
\eea
In contrast to the temperature in an equilibrium problem, the low-frequency effective temperature here 
is not a global property of the system, but is in general observable dependent. In Sec.~\ref{subsubsec:eff_temp}, we present a systematic, generalizable way to extract low-frequency effective temperatures based on 
observable-dependent fluctuation-dissipation relations. 
We already here quote the effective temperature for the atoms (see Sec.~\ref{sec:spontem} for the details of the computation):
\begin{align}
\widetilde{T}_{\phi}^{\text{eff}}=\frac{\gamma^2+\w_z^2}{4\w_z}\;,
\label{eq:fdrphi_intro}
\end{align}
where $\omega_z$ is the recoil energy and $\gamma$ an effective single atom decay rate. Because $\widetilde{T}_{\phi}^{\text{eff}}\neq
\widetilde{T}_{x}^{\text{eff}}$, the different parts of the driven system thus do not equilibrate to each other and, although the dominant low-frequency correlations are thermal, the system is not in a global thermal state. We remark that our definition of effective temperature does not coincide with the one commonly used in laser theory, as discussed in Sec.~\ref{subsubsec:eff_temp}: the former relates the fluctuations of the field to its response in the rotating frame, while the latter compares the fluctuations of the field to an equilibrium situation in the lab frame.

%

\subsection{Photon flux exponent}

The Langevin equation (\ref{eq:langevin_misha}) becomes dynamically unstable at the Dicke transition, correspondent to the point where $\alpha$ vanishes, or equivalently to the critical coupling
\begin{align}
g_c = \sqrt{\frac{\w_0^2+\kappa^2}{4\w_0}\w_z}\;.
\label{eq:g_c_misha}
\end{align}

Upon approaching the Dicke transition, the number of photons diverges as $|g-g_c|^{-\nu_x}$, where $\nu_x$ is called the ``photon flux exponent'' \cite{nagy11}. For the present non-equilbrium Dicke transition, it was found \cite{nagy11,oeztop11} that $\nu_x=1$, in contrast to the equilibrium case of a quantum phase transition at zero temperature\cite{emary03}, where $\nu_x=1/2$. We now explain that this discrepancy is due to the finite effective temperature of the low-frequency fluctuations of the system. 

The photon number is related to the fluctuations of $x$ by
\begin{align} 2\av{n}+1 &=2 \omega_0 \Big(\langle x^2 \rangle + \av{p^2}\Big) 
= 2\w_0\left(1+\frac{\kappa^2}{\w_0^2}\right)\av{x^2}, 
\label{eq:n}
\end{align}
Here we defined $p=i(\hat{a}-\hat{a}\yd)/\sqrt{2\w_0}$ and, by repeating the above derivation of the Langevin equation, observed that $\av{p}^2=(\kappa^2/\w_0^2)\av{x}^2$. We can compute $\av{x^2}$ using an equilibrium partition function equivalent \cite{einstein05,von_s06} to the (low-frequency limit) of the Langevin equation (\ref{eq:langevin_misha}): 
\be Z = \exp\left(-\frac{F}{T^{\rm eff}_{x}}\right),\quad{\rm with}\quad F=\half \a^2 x^2\;.\ee 
Performing the Gaussian integral we obtain
\begin{align}
2\av{n}+1 &= 2\left(1+\frac{\kappa^2}{\w_0^2}\right)\w_0\frac{T^{\rm eff}_{x}}{\a^2} 
\sim\frac1{|g-g_c|}\label{eq:infiniteN}\;, 
\end{align}
leading to the correct photon flux exponent $\nu_x = 1$.

The above results indicate that, from the point of view of phase transitions, it is incorrect to call the driven Dicke transition a quantum phase transition even though 
it is ``made of quantum ingredients'' (two collective motional states of a BEC \cite{baumann10}). Instead, it should be regarded as a classical phase transition belonging to the dynamical universality class of the classical Ising model with no conserved quantities and infinite-range interactions, a mean-field version of the ``Model A'' of Hohenberg and Halperin \cite{hohenberg77}. The effect that dissipation induces a finite effective temperature is not new. In several other condensed matter systems \cite{mitra06,MitraMillis,DallaTorre}, the coupling to a non-equilibrium bath typically admixes the pure many-body states of the 
closed system, transforming pure quantum phase transitions into 
thermal phase transitions (see also \cite{szymanska06,szymanska07,schwuehl08,diehl10,kessler12}). 
What is perhaps more surprising is that neither the low-frequency effective temperature for the photons, nor 
for the atoms, is set by 
the cavity loss rate $\kappa$, but instead set by the atom-photon interaction $g$ and the effective single atom parameters, respectively.

As a side remark, we note that, being complex-valued objects, the photons have two normal modes. One quadrature is thermally amplified and diverges at the Dicke transition. Its orthogonal quadrature is quantum squeezed \cite{dimer07}, remains gapped at the transition and therefore does not influence the thermal nature of the phase transition. These attenuated and amplified quadratures arise naturally as the eigenmodes of the photon correlation function (see Sec.~\ref{subsec:photoncorr}).

\subsection{Dynamic critical exponent}

In addition to the photon flux exponent, we identify a second indicator of criticality, the dynamical exponent. This exponent governs the decay of the two-time correlations close to criticality. 
Going back to the Langevin equation (\ref{eq:langevin_misha}) and keeping the $\w^2$ terms we obtain:
\begin{align}
&\left(V\w^2 + 2i\kappa \omega+\alpha^2\right) x(\w) = f(\w),\\
&\av{f(\w)f(\w')} = \kappa\frac{\w_0^2+\kappa^2+\omega^2}{\w_0}\delta(\w+\w')\;,
\end{align}
where we defined a dimensionless parameter $V = 1 + 4\omega_0g^2/\omega_z^3$. For simplicity we further approximate $V\approx 1$ 
and obtain the correlation function \footnote{In the vicinity of the phase transition the approximation $V\approx 1$ is justified only if $\w_0^2+\kappa^2\ll\w_z^2$.  However both the qualitative behavior of the correlation function and the analytic expression for the long-time asymptotics remain the same even beyond this limit.}
\begin{align}\label{eq:fullcorr}
 \langle\{ \hat x(t),\hat x(0) \}\rangle &= \mathrm i  G^K_{xx}(t)
 \\
  &=  \mathrm i \int \frac{d\w}{2\pi} ~e^{i\w t}\frac{\kappa (\omega_0^2+\kappa^2+\omega^2)}{2\omega_0[(2\kappa \omega )^2+(\alpha^2-\omega^2)^2] }\nn\\
&= \frac{e^{-t \kappa }}{8 \omega_0 m^2 \, \alpha^2}\left[m^2 \left( \omega_0^2 + \kappa^2 + \alpha^2\right) \text{cos}\left(\sqrt{m^2} t\right)\right.\nonumber\\
&\qquad\qquad\quad+\left.\kappa  \sqrt{m^2} \left( \omega_0^2 + \kappa^2 - \alpha^2\right) \text{sin}\left(\sqrt{m^2} t\right)\right]\nn\;,
\end{align}
where we defined $m^2 =  \alpha^2 - \kappa^2$. In the vicinity of the transition, for $0<\alpha<\kappa$, the frequency $m$ becomes purely imaginary and the oscillatory behavior in the above expression disappears. As already mentioned, this is a generic feature of dissipative phase transitions (see App.~\ref{app:natoverdamp}). For sufficiently large times and approaching the transition $\alpha\to 0$ (where only the closest pole to zero contributes), we then obtain
\begin{eqnarray}\label{eq:fullcorr2}
\langle\{ \hat x(t),\hat x(0)\} \rangle  = \tfrac{\omega^2_0 + \kappa^2}{8\omega_0 \alpha^2 } \,e^{-t /\xi_t}.
\end{eqnarray}
The correlation time:
\begin{eqnarray}
\xi_t  = \frac{2\kappa }{\a^2}  \sim \frac{1}{|g_c - g|^{\nu_t}}
\end{eqnarray}
is governed by a dynamical exponent $\nu_t = 1$.

The remainder of the paper is dedicated to the development of 
a unified, and generalizable, Keldysh approach. The Dicke model will be used as a prototypical test object and we 
compare our results to those of other approaches, where available. 


\section{Keldysh approach for cavity vacuum}
\label{sec:cavityonly}

In this section, we introduce the real-time Keldysh formalism and fix our notation by considering the case of a single 
electromagnetic mode in an open cavity (without atoms). We also indicate 
how to relate the operators of the master equation (\ref{eq:master-gen}) to the field content and choice of time contour 
in a Keldysh action (see also Refs.~\cite{castroneto90,takei08,sarang10}).

The decay of a single boson (the cavity photon) into a continuum of modes (the external vacuum) is described by the master equation
\be \partial_t \rho = -\mathrm i [\w_0 \hat a\yd \hat a , \rho] + \kappa(2 \hat a \rho \hat a^\dag - \{\hat a^\dag \hat a , \rho\} ).
\label{eq:master0} \ee
This equation results from the (fully unitary) Heisenberg equation for the coupled system-bath setting, where the system is described by the degrees of freedom $\hat a,\hat a^\dag$ and the bath by a continuum of harmonic modes. Eq. \eqref{eq:master0} is obtained by eliminating (``integrating-out'') the bath in the Born-Markov and rotating wave approximation  (cf. e.g. \cite{zollerbook}). The integration of the bath variables gives rise to an effective evolution including dissipative terms. Performing the same program in path integral formulation, we obtain the \emph{Markovian dissipative action}
\begin{eqnarray}\label{eq:Scavity}
 S_a &=&  \int_{-\infty}^\infty d t \Big( a^*_{+}(\mathrm i \partial_t  - \omega_0) a_{+}  -  a^*_{-}(\mathrm i \partial_t  - \omega_0) a_{-}  \\\nonumber
 &&\quad \quad\,\,- \mathrm i \kappa [ 2 a_{+} a_{-}^*  -( a_{+}^* a_{+} +a_{-}^* a_{-}  )]\Big).
\end{eqnarray}
In the path integral formalism, the quantum mechanical operators are replaced by fluctuating, time-dependent and complex-valued fields (we omit the time argument for notational simplicity).  The fact that the density matrix can be acted on from both sides, as reflected in the Heisenberg commutator structure of the original evolution equation, finds its counterpart in the presence 
of a forward (+) and backward (-) component of the fields.
The former is associated to an action on the density matrix from the left, and the latter to the right. Indeed, in the first of line of 
Eq. (\ref{eq:Scavity}) 
there is a relative minus sign between the terms involving the two components, reflecting the Heisenberg commutator structure of Eq.~(\ref{eq:master-gen}). The terms in the second line instead display the characteristic Lindblad form; the ``jump'' or ``recycling'' term 
is represented by an explicit coupling of the two contours.

It is  convenient \cite{kamenev09,kamenev_book,altlandbook} to introduce ``center-of-mass'' and ``relative'' field coordinates, $a_{cl} = (a_+ + a_-)/\sqrt{2}, a_{q} = (a_+ - a_-)/\sqrt{2}$. These new coordinates are often referred to as ``classical'' and ``quantum'' fields, because the first can acquire an expectation value while the second one cannot in the absence of sources. In this basis, and going 
to frequency space, we write
\begin{eqnarray}\label{eq:oscRAKaction}
S_a&=& \int_{\omega} (a^*_{cl}, a^*_{q}) 
\left(\begin{array}{cc}
0 &  {[G^{A}]}^{-1}(\w) \\ {[G^{R}]}^{-1}(\w)  & D^K(\w)
\end{array}\right)
\left(\begin{array}{c}
a_{cl}\\ 
a_{q}
\end{array}\right)\;,
\end{eqnarray}
where we used the notation $\int_{\omega}=\int_{-\infty}^{\infty}\frac{d\omega}{2\pi}$, and $a_{cl,q}(t) = \int_{\omega} e^{-i\omega t} a_{cl,q}(\omega)$.
This classical-quantum basis is often referred to as RAK basis: the entries are the inverse Retarded (lower left) and Advanced (upper right) Green's functions, and the inverse Keldysh component. The RAK action Eq. (\ref{eq:oscRAKaction}) can be easily inverted to deliver the photonic Green's functions
\bea\label{eq:GFtotal}
\left(\begin{array}{cc}
G^K(\omega)   & G^R(\omega) \\
 G^A(\omega) &  0
\end{array}\right) = \left(\begin{array}{cc}
0 & [G^{A}]^{-1}(\omega) \\
{[G^{R}]}^{-1}(\omega) &  D^K(\omega)
\end{array}\right)^{-1}\;,
\nonumber\\
\eea
where the Keldysh Green's functions is a matrix product 
\bea
G^K(\omega) &=& - G^R(\omega) D^{K}(\omega) G^A(\omega)\;.\label{eq:gk}
\end{eqnarray}
For the open cavity of Eq. (\ref{eq:Scavity}) the RAK inverse Green's functions are
\begin{align} 
{[G^{-1}]}^{R/A}& = \omega - \omega_0 + \Sigma^{R/A}_{a\text{\MARKOV}}\;,\quad
D^K  = \Sigma^{\text{K}}_{a\text{\MARKOV}},\label{eq:oscGRAK}
\end{align} 
with the ``self-energies''
\begin{align}
\Sigma^{\text{A}}_{a\text{\MARKOV}}&=-i\kappa,\quad
\Sigma^{\text{R}}_{a\text{\MARKOV}}=+i\kappa,\quad
\Sigma^{\text{K}}_{a\text{\MARKOV}}=2 i \kappa\,\,.\label{eq:photon_self_markov}
\end{align}
It is a key property of a Markovian system that the Keldysh component  $\Sigma^{\text{K}}_{a,\text{\MARKOV}}$ in Eq.~(\ref{eq:photon_self_markov}) is frequency independent. As can be seen from App.~\ref{app:integration}, 
this is due to a separation of scales 
between (i) the large pump ($\omega_p$) and cavity frequency ($\omega_c$), both optical frequencies in 
the Tera Hertz range ($\sim10^{14}$ Hz corresponding to temperatures $T \sim 10^4 K$), and (ii) the characteristic 
frequencies of the electromagnetic vacuum outside the cavity ($\lesssim 10^{12}$ Hz corresponding to temperatures 
$T\lesssim 300 K$). 

In the literature, it is often argued that a frequency independent, nonzero inverse Keldysh component indicates an effective finite temperature state. 
The Markovian lossy cavity is a simple counterexample: Even though the inverse Keldysh component is constant $\sim 2i\kappa$, the state is pure and the effective temperature zero, as we will argue below.

We next introduce the key propagators that encode the systems' {\em response} and {\em correlation} functions. In equilibrium, 
the two are rigidly related by the Bose (or Fermi) distribution function; out-of-equilibrium, no such \emph{a priori} knowledge is available, and it is important to distinguish them.

\subsection{Cavity spectral response function}
\label{subsec:spectral}
The {\em spectral response function} encodes the system's
response to active, external perturbations such as time-modulated external fields coupling to spin or charge operators, for example. 
The spectral response function is the difference between the retarded and advanced Green's function:
\begin{eqnarray}\label{eq:spectralF}
\mathcal A (\w) = \mathrm i (G^R (\w)  - G^A(\w) ).
\end{eqnarray}
In the scalar case considered here, we have
\begin{eqnarray}
\mathcal A_{aa\yd} (\w) = - 2 \mathrm{Im} G^R(\w),
\label{eq:Aaa}
\end{eqnarray}
The frequency-integrated spectral response function is normalized to unity, because of the exact commutator relation of the bosonic 
degrees of freedom 
\be
\int \frac{d \w}{2\pi} \mathcal A_{aa\yd}(\omega)  = \langle [  \hat a, \hat a^\dag]\rangle = 1\;.
\ee
This ``sum rule'' is an exact property of the theory valid in-- and out--of-- equilibrium.
In our example of one cavity mode,
\be
\mathcal A_{aa\yd} (\w) = \frac{2\kappa}{(\omega - \omega_0)^2 + \kappa^2}\;.
\label{eq:cav_response}
\ee
%
%
\subsection{Cavity correlation function}
\label{subsec:correlation}
The {\em correlation function} encodes the system's internal 
correlations, for example the frequency-resolved photon spectrum of the intracavity photon fields.
In the steady state, the photon correlation function is related to the Keldysh Green's function by
\begin{align}
\mathcal C_{aa\yd} (t) =  \langle\{ \hat a(t),\hat a^\dag(0)\}\rangle =  \av{\hat a(t)\hat a^\dag(0)+ \hat a^\dag(0)\hat a(t)}
= \mathrm i G^K(t).\label{eq:Caa}
\end{align}
Here the last identity is valid only in the specific case of a scalar Keldysh Green's function. At equal times this relation results in 
\bea \label{eq:anticomm}
\mathcal C_{aa\yd} (0) = 2\langle \hat a^\dag \hat a \rangle +1 = \mathrm i G^K(t=0)=\mathrm i   \int \frac{d \w}{2\pi} G^K(\omega)\;.
\eea
For the single decaying cavity mode, characterized by Eqs.~(\ref{eq:oscRAKaction}-\ref{eq:photon_self_markov}), it is easy to show that
$\mathcal{C}_{aa^\dagger}(\omega)=\mathcal{A}_{aa^\dagger}(\omega)$ and 
the frequency integral over the Keldysh Green's function is unity yielding $\av{\hat a\yd \hat a}=0$. 
As expected, the steady state corresponds to the cavity vacuum.
\subsection{Comparison with a closed system at equilibrium}
\label{subsec:comparison}
In the absence of dissipation, the cavity becomes an isolated harmonic oscillator. Its inverse Green's functions are still given by (\ref{eq:oscRAKaction}) with the self-energies serving only as regularization parameters. At equilibrium,
\begin{align}
\Sigma^{\text{A}}_{a,\text{\EQUILIB}}&=-i\epsilon, \nonumber\quad
\Sigma^{\text{R}}_{a,\text{\EQUILIB}}=+i\epsilon,\nonumber\\
\Sigma^{\text{K}}_{a,\text{\EQUILIB}}(\omega)&=2 i \epsilon \coth\left[\frac{\omega}{2T}\right]\,\,.
\label{eq:photon_self_equilib}
\end{align}
Here $\epsilon\rightarrow 0$ at the end of the calculation and $T$ is the actual temperature. Note that the Keldysh component is odd with respect to the frequency $\Sigma^{\text{K}}_{a,\text{\EQUILIB}}(-\w) = - \Sigma^{\text{K}}_{a,\text{\EQUILIB}}(\w)$, while in the Markovian system (\ref{eq:photon_self_markov}) it is even. 

Using Eq.~(\ref{eq:photon_self_equilib}), we obtain
\begin{eqnarray}
\mathcal A^{\text{\EQUILIB}}_{aa\yd}(\w) &=& \lim_{\epsilon\to 0}\frac{2\epsilon}{(\omega - \omega_0)^2 + \epsilon^2} = 2\pi \delta (\omega - \omega_0),\\ \mathcal{C}^{\text{\EQUILIB}}_{aa\yd}(\omega)&=& \lim_{\epsilon\to 0}\frac{2 \mathrm  \epsilon}{(\omega - \omega_0)^2 + \epsilon^2}\coth \tfrac{\omega}{2T}  = 2\pi\mathrm \coth \tfrac{\omega}{2T} \delta (\omega - \omega_0), \nonumber
\end{eqnarray}
In this noninteracting case, the spectral response function is fully centered at the isolated mode with frequency $\omega_0$. We observe that formally, the thermodynamic equilibrium limit can be seen as a situation with an infinitesimal loss, replacing $\kappa \to \epsilon$, and the replacement in the inverse Keldysh component $2 \mathrm i \kappa \to 2\mathrm i \epsilon \coth  \tfrac{\omega}{2T}$.

\subsection{Cavity distribution function and low-frequency effective temperature}
\label{subsec:eff_temp}
The response and correlations allow us to define a fluctuation-dissipation relation, by introducing  the {\em distribution function} $F(\omega)$:
\begin{eqnarray}
G^K(\w) &=& G^R(\w) F(\w) - F(\w) G^A(\w)\label{eq:defineF}\\\nonumber
\Leftrightarrow D^K(\omega ) &=&{ [G^{R}(\w)]}^{-1} F(\w) - F(\w) {[G^{A}(\w)]}^{-1},
\end{eqnarray}
where the equivalence holds due to Eq. \eqref{eq:gk}. At thermal equilibrium the distribution $F$ is universal and  equals to the unit matrix times
\bea
F^{\text{\EQUILIB}}(\w) &=&  \coth \tfrac{\omega}{2T}  =  2 n_B (\tfrac{\omega}{T}) +1,\nonumber\\
F_{T=0}^{\text{\EQUILIB}}(\w)&=&{\rm sign}(\w),\nonumber\\
F_{\omega \ll T }^{\text{\EQUILIB}}(\w)&\approx& \frac{2 T}{\omega}+...
\label{eq:EQ_F}
\eea
with the Bose distribution $n_B (x) = (\exp x - 1)^{-1}$. The unit matrix in field space signals detailed balance between all subparts of the system.

In the present case, the system is out-of-equilibrium due to its driven and dissipative nature. A notion of a temperature is not \emph{a priori} meaningful: Neither must the driven system equilibrate to an external heat bath with temperature $T$ (in our case, due to the separation of scales underlying the Markov approximation, this temperature would be effectively $T=0$ compared to the system scales), nor do the different subparts of the system have to equilibrate with respect to each other. In this work, we argue that a notion of a temperature nevertheless emerges as a universal feature of the low frequency domain of Markovian systems. It is introduced by computing the $F$ matrix through Eq.~(\ref{eq:defineF}) and comparing the low-frequency behavior of its eigenvalues with the equilibrium result of Eq. (\ref{eq:EQ_F}). In particular, if $F$ has a thermal infrared enhancement $\sim1/\omega$ for small frequencies, its dimensionful coefficient is identified as an effective temperature. This notion of a ``low frequency effective temperature'' (LET) becomes particularly relevant in the vicinity of a phase transition, where the spectral weight encoded in $G^R,G^A$ is concentrated near zero frequency. Below, we will use this concept to establish a connection of Markovian quantum systems to the classical theory of dynamical universality classes according to Hohenberg and Halperin \cite{hohenberg77}. Moreover, we find that, while all governed by the $1/\omega$ divergence in the distribution function, different subparts of the system exhibit different LETs. In contrast to the global temperature present in thermodynamic equilibrium, the LET is not an external parameter but rather a system immanent quantity, determined by the interplay of unitary and dissipative dynamics. 


In case of a decaying cavity, the Green's functions are scalars and we can easily invert (\ref{eq:defineF}) to obtain
\be 
F(\w) = \frac{G^K(\w)}{G^R(\w) - G^A(\w)} = \frac{\mathcal{C}_{aa\yd}(\omega)}{\mathcal A_{aa\yd}(\omega)} = 1.
\label{eq:F_ratio}
\ee
confirming that the cavity vacuum has a zero effective temperature.   Moreover, 
as for a pure quantum state in the equilibrium case at $T=0$, the distribution function here also 
squares to a unit matrix, $F^2(\omega) =1$.

An important difference between the zero temperature equilibrium case and Markovian case appears in the sign of the distribution function. In the former case (as for any equilibrium distribution), $F(\w)$ is anti-symmetric with respect to the frequency $\omega$. On the contrary, for a Markovian bath the distribution function is symmetric with respect to $\w$; one signature of a strongly-out-of-equilibrium system.



\section{Keldysh approach for photon observables}
\label{sec:dicke_kappa}

We now analyze the Dicke model Eq.~(\ref{eq:HDicke1}) with the path integral approach 
explained in the previous section. We include cavity photon loss but defer the inclusion of dissipative processes for the atoms 
to Sec.~\ref{sec:spontem}. 
Assuming homogeneous qubit-cavity coupling, one can use the large-$N$ strategy of Emary and Brandes \cite{emary03}. 
We introduce collective large-$N$ spin operators $S_z=\half\sum_{i=1}^N \sigma^z_i$ and $S^x =\half\sum_{i=1}^N (\sigma^+_i +\sigma^-_i)$ ,
to write the Dicke model (\ref{eq:HDicke1}) in terms of one large spin coupled to the cavity photon mode,
\be 
H = \w_0 \hat a^\dag \hat a  + \w_z S_z + \frac{2g}{\sqrt{N}} S^x \left(\hat a\yd +\hat a\right) \label{eq:Hhomo} \;.
\ee
We then express the spin in terms of a Holstein-Primakoff \cite{holstein40,emary03} boson operator $\hat b$, defined by
$S_z = - N/2 + \hat b\yd \hat b$, $S^+ = \sqrt{N-\hat n}\hat b\yd \approx \sqrt{N}(1+\hat n/(2N))\hat b^\dag $, and $S^x=(S^++S^-)/2$. Neglecting unimportant constants we obtain the normal ordered Hamiltonian:
\begin{eqnarray} \label{eq:large_N_hamil}
H &=& \w_0 \hat a^\dag \hat a + \w_z \hat b\yd \hat b \\\nonumber
&& + g \left(\hat a+ \hat a^\dag \right) \left(\hat b + \hat b\yd - \tfrac{1}{2N}\hat b\yd (\hat b\yd+\hat b) \hat b\right).
 \end{eqnarray}
At $N\rightarrow \infty$, the last, non-quadratic term vanishes and the problem reduces to a linear system of two coupled bosonic degrees of freedom one of which (the cavity mode $\hat a$) decays into a Markovian bath.
As outlined in section \ref{sec:cavityonly}, we can transform the Liouvillian Eq.~(\ref{eq:cav_liou}), with Hamiltonian Eq.~(\ref{eq:large_N_hamil}), into an equivalent Keldysh action  with
\begin{align}
\label{eq:SDicke}
S &= S_a + S_b + S_{ab}\\
S_b &= \int_\omega (b^*_{cl} , b^*_{q}) 
\left(\begin{array}{cc}
0 & \mathrm \omega  - \omega_z \\
 \mathrm \omega   - \omega_z &0
\end{array}\right)
\left(\begin{array}{c}
b_{cl}\\ 
b_{q}
\end{array}\right)\nonumber
\end{align}
where $S_a$ is given by Eq.~(\ref{eq:oscRAKaction}) and 
the interaction in terms of the ``classical'' and ``quantum'' fields reads
%
\begin{align}
&S_{ab}= - g \int_\omega [(a_q + a_q^*)(b_{cl} + b_{cl}^*) + (a_{cl} + a_{cl}^*)(b_{q} + b_{q}^*)]
\nonumber\\
&-\frac{1}{4N}\Bigg\{\left[(a_{cl} + a_{cl}^*)(b_{q} + b_{q}^*) + (a_{q} + a_{q}^*)(b_{cl} + b_{cl}^*)\right]
\left[b_{cl}^* b_{cl} + b_{q}^* b_{q}\right]\nn \\
&+\left[(a_{cl} + a_{cl}^*)(b_{cl} + b_{cl}^*) + (a_{q} + a_{q}^*)(b_{q} + b_{q}^*)\right]\left[b_{cl}^* b_{q} + b_{q}^* b_{cl}\right] \Bigg\}\;.
\label{eq:Squartic}
\end{align}
%
We now demonstrate that the static saddle point solutions of this action reproduce the results 
of other approaches \cite{dimer07}. Varying $S$ with respect to the quantum components of the fields and substituting $a_{cl}(t)=\sqrt{2}a_0,~b_{cl}(t)=\sqrt{2}b_0,~a_q=0,~b_q=0$, we obtain the coupled equations
\bea 
\frac{\partial S}{\partial a_q^*} &=& (-\w_0+i\kappa)a_0 - g \left(1 - \frac{1}{2N} b^2_0\right)2b_0 = 0\;,\label{eq:saddlepoint1}\\
\frac{\partial S}{\partial b_q^*} &=& -\w_z b_0 - g \left(1 - \frac{3}{2N} b^2_0\right)(a_0+a^*_0) = 0\;. \label{eq:saddlepoint2}
\eea
where we chose $b_0=b_0^*$. 
These saddle-point equations admit solutions with non-zero ``ferromagnetic'' moment 
$b_0$, and super-radiant photon condensate $a_0$,
%
%
%
\be
b_0 = \pm \sqrt{\frac{N}2}\sqrt{\frac{g^2-g_c^2}{g^2}}~~~a_0 =  \pm \sqrt{2N}\frac{\sqrt{g^2-g_c^2}}{\w_0-i\kappa}\label{eq:psi0}
\ee
for atom-photon couplings larger than a critical value 
\be 
g > g_c = \sqrt{\frac{\w_0^2+\kappa^2}{4\w_0}\w_z}\;,
\label{eq:g_c_keldysh}
\ee
in agreement with the results known from the literature\cite{nagy11,oeztop11}. We are now going to integrate-out the atomic field $b$ and obtain an effective action describing the photons in the normal phase ($g<g_c$), where $a_0=b_0=0$. 
In App.~\ref{app:superradiant}, we give the corresponding expressions in the superradiant phase.

In the limit of $N\to\infty$, we can safely neglect the terms proportional to $1/N$ in (\ref{eq:SDicke}). 
We can re-write Eq.~(\ref{eq:SDicke}) as $8\times8$ matrix multiplying the 8-component fields: 
\begin{align}
V_8(\omega) =\left(\begin{array}{l} 
a_{cl}(\w)\\
a_{cl}^* (-\w)\\
b_{cl}(\w)\\
b_{cl}^* (-\w)\\
a_{q}(\w)\\
a_{q}^* (-\w)\\
b_{q}(\w)\\
b_{q}^* (-\w)
\end{array}\right)
\label{eq:V_8}
\end{align}
with the action
\begin{eqnarray}\label{eq:Snormal} 
S_{\rm N} =\frac{1}{2}\int_{\omega} V_{8}^\dag(\omega)
\left(\begin{array}{cc}
0 & {[G_{4\times4}^A]}^{-1}(\omega) \\
 {[G_{4\times 4}^R]}^{-1}(\omega) &  D_{4 \times 4}^K
\end{array}\right)V_{8}(\omega)\;.
\nonumber\\
\end{eqnarray}
The subscript $\rm N$ stands for normal phase and the dagger $\dagger$ denotes transposition 
and complex conjugation.
The block entries are $4\times 4$ Green's functions given by
\begin{align}\nonumber
{[G_{4\times 4}^R]}^{-1}(\w) &=  \left(\begin{array}{cccc}
 \omega - \omega_0 + i \kappa & 0  & -g & -g \\
0 & - \omega - \omega_0 - i \kappa & -g & -g \\
-g & -g &\omega - \omega_z   & 0  \\
-g & -g & 0&- \omega - \omega_z   \\
\end{array}\right)\nn
\\\
D_{4\times4}^{K}&= 2 \mathrm i \,\,\mathrm{diag} (\kappa,\kappa,0,0).\label{eq:G4x4}
\end{align}
To obtain the photon-only action, we now integrate out the Holstein-Primakoff 
field $b$ and get a Keldysh functional integral that goes only over the photon fields:
%
$Z^{\text{K}}=\int D \{a^*,a\} e^{i S_{\text{photon}}[a^*,a]}$.
%
with the photon-only action
\begin{eqnarray}
S_{\text{photon}} [a^\ast,a]&=& \int_{\omega} A_4\yd(\w)\left(\begin{array}{cc}
0 &  {[G_{2\times 2}^{A}]}^{-1}(\w) \\ {[G_{2\times 2}^{R}]}^{-1}(\w)  & D_{2\times 2}^K(\w)\ea\right)
A_4(\w)\;.\nonumber\\
\label{eq:Sphotononly}
\end{eqnarray}
The photon four-vector collects the classical and quantum field components
\begin{align}
A_4(\omega) =\left(\begin{array}{l} 
a_{cl}(\w)\\
a_{cl}^* (-\w)\\
a_{q}(\w)\\
a_{q}^* (-\w)
\end{array}\right)\; .
\end{align}
and the block entries are $2\times2$ photon Green's functions which we now analyze one-by-one.

\subsection{Photon spectral response function}
The inverse retarded Green's function of the photons is
\begin{align}
&{[G^{R}_{2\times2}]}^{-1}(\omega) =\nn\\&  \left(\ba{c c} \omega-\w_0 + i\kappa +\Sigma^R(\w) &\Sigma^R(\w) 
\\ {\left[\Sigma^R(-\w)\right]}^*  &-\omega-\w_0 - i\kappa + {\left[\Sigma^R(-\w)\right]}^* \ea\right)\;,
\label{eq:GRphoton}
\nonumber\\
\end{align}
where the interaction induced photon-self energy reads
\be \Sigma^R(\w) =  -\frac{2g^2\w_z}{\omega^2-\w^2_z}\;.
 \ee
\begin{figure}[t]
\includegraphics[scale=0.7]{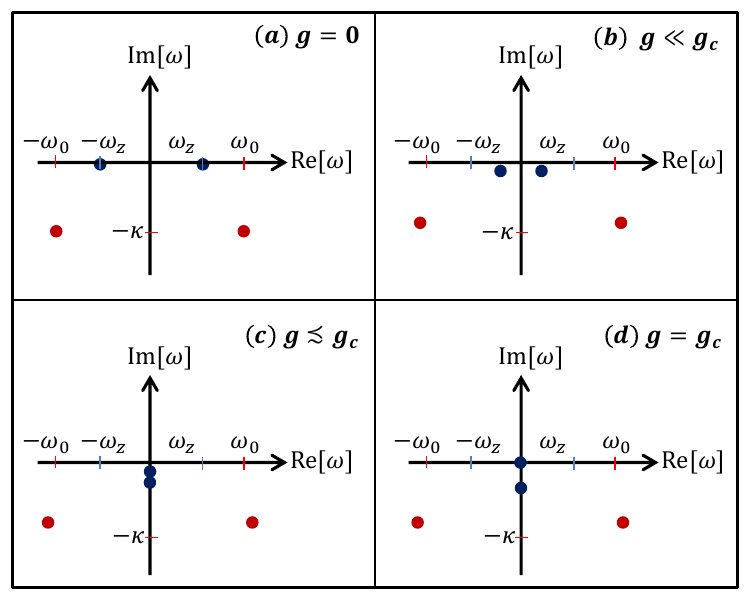}\\[5mm]
\caption{(Color online) Schematic plot of the position of the poles of the retarded Green's function Eq. (\ref{eq:eigen}). (a) At zero coupling $g = 0$, two poles can be associated with the photonic mode $\w = \pm \w_0 +i\kappa$ and two with the atomic mode $\w = \pm \w_z$. (b) In the presence of a finite coupling $g$, the modes hybridize and the corresponding frequencies are shifted in opposite directions. (c) When approaching the transition, two solutions become purely imaginary and correspond to damped modes. (d) At the transition point $g=g_c$ one of the poles approaches zero, making the system dynamically unstable.} 
\label{fig:poles}
\end{figure}
The characteristic frequencies of the system are defined by the zeros of the determinant ${[G^{R}_{2\times2}]}^{-1}(\omega)$, corresponding to the poles of the response function $G^R_{2\times2}(\w)$. Due to the symmetry 
\begin{eqnarray}\label{eq:symm}
\sigma_x {\left[G_{2\times2}^{R }(-\omega)\right]}^* \sigma_x = G^R_{2\times2}(\omega)\;,
\end{eqnarray}
the poles come in pairs, such that $\{\lambda\}=\{-\lambda^*\}$, meaning that they either are pure imaginary or come in pairs with opposite real part. The explicit solution of 
\be
0 = \frac1{\mathrm{det}[G^{R}_{2\times2}(\omega)]} = \left(\w_0 + \frac{2g^2\w_z}{\omega^2-\w_z^2}\right)^2 -  (\omega + i\kappa)^2 -  \left(\frac{2g^2\w_z}{\omega^2-\w_z^2}\right)^2 \label{eq:eigen} \ee
yields four poles, schematically plotted in Fig. \ref{fig:poles}. 
Note that, in the vicinity of the phase transition, two poles become purely imaginary. This phenomenon seems to apply to generic dissipative transitions  as we further describe in App.~\ref{app:natoverdamp}. Indeed it has been previously  observed 
for a dissipative critical central spin model \cite{kessler12}. Overdamping of collective modes, due to a similar mechanism, has also been found in 
dissipative multimode systems in symmetry broken phases \cite{szymanska06,szymanska07}. 

\begin{figure}[t]
\includegraphics[scale=1]{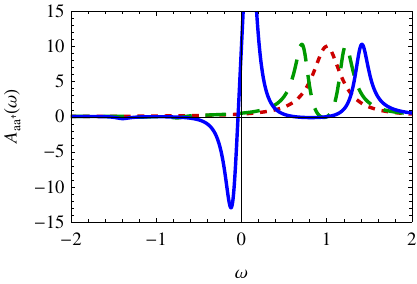}\\[5mm]
\includegraphics[scale=1]{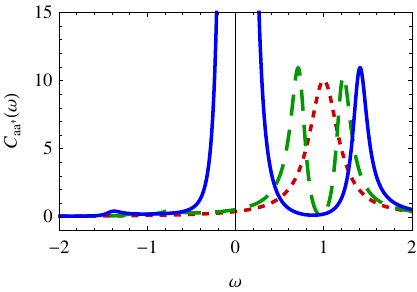}
\caption{(Color online) Photon spectral response function $A_{aa\yd}(\w)$ and correlation function $C_{aa\yd}(\w)$ 
as a function of real frequencies
$\omega$. Numerical parameters: $\w_0=\w_z=1,~\kappa=0.2$ (leading to $g_c\approx0.51$), and g=0 (dotted), 0.25 (dashed), 0.5 (solid). }
\label{fig:photon_response}
\end{figure}

The imaginary part of the first diagonal element of $G^R_{2\times2}(\w)$ corresponds to the photon spectral response function $\mathcal{A}_{aa\yd}$, defined in (\ref{eq:spectralF}), and is plotted in Fig.~\ref{fig:photon_response} for different values of the coupling $g$. In the absence of atom-photon coupling $g=0$ (dotted curve) 
there is a single resonance peak at frequency $\w=\w_0$, broadened by the cavity decay rate $\kappa$. A finite coupling $g$ (dashed curve) ``collectively Rabi-splits'' the resonance \cite{thompson92} in two distinct peaks, corresponding to two distinct poles of the system. Upon approaching the Dicke transition (solid curve), the spectral weight is shifted towards the low-frequency pole; a precursor to the superradiant cavity mode.

%

\subsection{Photon correlation function}
\label{subsec:photoncorr}
The inverse Keldysh component of the action Eq.~(\ref{eq:Sphotononly}) is
\be\label{eq:GKphoton}
D_{2\times2}^K = \left(\begin{array}{cc}
2i\kappa & 0 \\ 0 & 2i\kappa
\end{array}\right)
\ee
and the Keldysh Green's function $G_{2\times2}^K(\omega) = -G^R_{2\times2}(\omega) D^K_{2\times2} G^A_{2\times2}(\omega)$ is a $2\times2$ matrix.

The first diagonal element of $G^K_{2\times2}$, $\mathcal C_{aa\yd} = \rm{i} (1 ~0)G_{2\times2}^K(1~0)^T$ corresponds to the photon correlation function defined in Eq.~(\ref{eq:Caa}) and is plotted in Fig. \ref{fig:photon_response}. As noted above in Eq.~(\ref{eq:anticomm}), its frequency integral  gives the steady-state photonic occupation and it can be shown to diverge at the Dicke transition according to Eq.~(\ref{eq:n}). This result will be explicitly derived in Sec. \ref{subsec:langevin_keldysh} using a low-frequency effective description of $G^K$.
The photon number diverges in steady state despite the fact that the system undergoes photon loss, and no explicit photon pumping occurs within the model. The reason is that the coupling constant $g$ is an effective parameter, which in any concrete physical realization microscopically involves a coherent laser drive process compensating for the loss. 

\begin{figure}[b]
\vspace*{-4mm}
\includegraphics[scale=1]{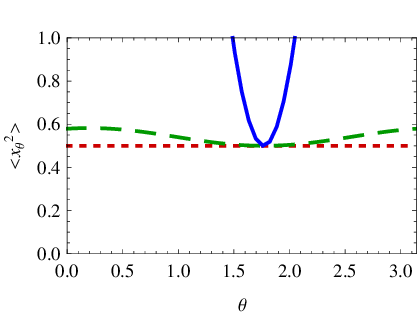}
\caption{Frequency integrated equal-time correlations $\av{\left(x_{\theta}(t)\right)^2}$ as function of the angle $\theta$, for different values of the coupling strength approaching the transition at $g_c\approx0.51$: g=0 (dotted), 0.25 (dashed), 0.5 (solid). The fluctuations in the quadrature $\theta^*=\pi-{\rm tan}^{-1}(\w_0/\kappa)\approx 1.768$ are independent of the coupling strength. Numerical parameters: $\w_0=\w_z=1.0$, $\kappa=0.2$}
\label{fig:photon_theta}
\end{figure}

The matrix structure of the Keldysh Green's function $G^K_{2\times2}(\w)$ can be conveniently exploited to compute the quadrature fluctuations
for a general phase angle $\theta$
\be\av{x_{\theta}(\w)x_\theta(-\w)} = \frac{\mathrm i}{4\w_0}\left(e^{-i\theta}~e^{i\theta}\right) G^K_{2\times2}(\w) \left(\ba{c}e^{i\theta}\\e^{-i\theta}\ea\right)
\label{eq:xthetaxtheta},\ee
where $x_\theta$ is defined by
\be 
x_{\theta} = \frac{1}{\sqrt{2\w_0}}\left(e^{i\theta}a + e^{-i\theta}a\yd\right).\label{eq:xtheta}
\ee
The corresponding equal time, frequency-integrated fluctuations $\av{x_\theta(t) x_\theta(t) }=\int_\w \av{x_{\theta}(\w)x_{\theta}(-\w)}$ are plotted in Fig.~\ref{fig:photon_theta} and diverge at the Dicke transition for all angles $\theta$ except for $\theta^\ast$ defined by 
\be \theta^*=\pi-{\rm tan}^{-1}(\w_0/\kappa)\;. \label{eq:thetastar}\ee
The angle $\theta^\ast$ can also be obtained as 
the phase angle for the non-diverging eigenmode of the zero-frequency limit of the Keldysh correlation function, thereby 
naturally yielding the attenuated and amplified quadratures alluded to in Sec.~\ref{sec:review}.

Here in the case of the driven Dicke model, the {\em equal-time photon fluctuations} of $x_\theta*$ inside the cavity are independent of the atom-photon coupling and in particular are not attenuated below the vacuum noise level (the $g=0$ limit, without atoms in the cavity). 
This is different from the case of the optical parametric oscillator \cite{collett84,mertens93} where, at threshold, the equal-time fluctuations of the 
non-diverging intra-cavity quadrature are reduced to 50\% of the vacuum level. This difference can be traced 
back the different frequency dependencies of the effective driving term in both situations. 
In the parametric oscillator, the driving of the cavity occurs via a classically-treated photon pump laser, and the driving amplitude 
is typically set to a constant coherent field amplitude.
In the present case of the driven Dicke model, the effective driving of the cavity is mediated by the atoms via virtual absorption and emission 
of photons from the pump laser into the cavity. The corresponding driving term $\sim g^2\w_z/(\w_z^2-\w^2)$ is maximal for frequencies $\omega$ of the order of the atomic detuning $\omega_z$ and vanishes for large frequencies.

Nevertheless, also in the driven Dicke model, the experimentally relevant homodyne spectrum $G^{\rm out}_{2\times2}(\w)$ of the 
cavity output field shows noise reduction below the vacuum level in the $\theta^*$ quadrature \cite{dimer07}. Following the standard  ``input-output theory'' \cite{collett84, carmichael_book}, it is possible to show that the homodyne spectrum can be linked to the Keldysh response function via

\be \mathrm{i}~G^{\rm out}_{2\times2}(\w) = \left|\left|\left(\ba{c c}2i\kappa&0\\0&-2i\kappa\ea\right)G^R_{2\times2	}(\w)-\mathbf{1}_{2\times2}~\right|\right|^2-\mathbf{1}_{2\times2},\ee
where $\left|\left|M\right|\right|^2 \equiv M\yd M$. By applying the transformation (\ref{eq:xthetaxtheta}) to $G^{\rm out}_{2\times2}$, it is then possible to compute the fluctuations of the output quadrature $\av{x_{\rm out,\theta}(\w)x_{\rm out,\theta}(-\w)}$. This quantity has been studied in detail in Ref.~\cite{dimer07}: at the Dicke transition, the zero-frequency component of $\theta=\theta^*$ tends to the maximally attenuated value of $-1/(4\w_0)$. It should be noted, however, that the equal-time, frequency integrated fluctuations $\av{x_{\rm out,\theta^*}(t)x_{\rm out,\theta^*}(t)}$  are zero and therefore not attenuated below vacuum level.

\subsection{Comparison with a closed system at equilibrium}
In the case of a closed system at equilibrium, the retarded Green's function is the same as Eq.~(\ref{eq:GRphoton}), up to the replacement $\kappa \to \e$, and letting $\e \to 0$. The corresponding spectral response function $\mathcal{A}_{aa\yd}(\w)$ is similar to the one shown for the Markovian case, with the narrow peaks substituted by delta-functions at the resonant frequencies. 
The Keldysh component of the inverse Green's function reads, at zero temperature:
\be
D^{K}_{2\times2,~{\rm \EQUILIB}}= \left(\begin{array}{cc}
2i\e{\rm sign}(\w) & 0 \\ 0 & -2i\e{\rm sign}(\w)
\end{array}\right).\label{eq:GKphotoneq}
\ee
Again note the different symmetry under frequency reflection with respect to (\ref{eq:GKphoton}). 

%

\subsection{Photon distribution function and low-frequency effective temperature}
\label{subsubsec:eff_temp}

\begin{figure}[t]
\includegraphics[scale=1]{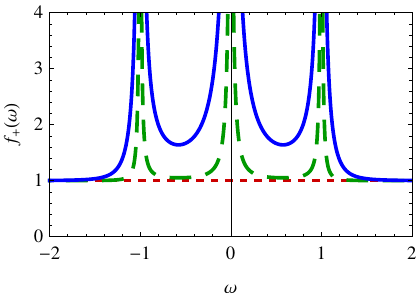}
\caption{(Color online) Positive eigenvalue of the distribution function $F(\w)$ for the same parameters as in Fig. \ref{fig:photon_response}. At low frequencies the distribution diverges as $2\widetilde{T}_{\rm eff}/\omega$, where the finite effective temperature is proportional to the photon-atom interaction, $\widetilde{T}_{\rm eff} \sim g^2$.} \label{fig:photon_distribution}
\end{figure}

We now show that the above mentioned difference between the Markovian case and the equilibrium case leads to the generation of a ``low frequency effective temperature'' (LET) for the former. For this purpose, we calculate the distribution matrix $F$, defined in Eq.~(\ref{eq:defineF}). To this end, recall that at thermal equilibrium, $F=\coth(\omega/2T)\mathbf 1$: it exponentially approaches unity at high frequencies ($|\omega| \gg T$), and diverges as $2T/\omega$ at low frequencies ($|\omega| \ll T$).  For our Markovian problem, using Eq.~(\ref{eq:Sphotononly}) we find:
\be F = \sigma_z + \frac{2}{\omega}\frac{g^2\w_z}{\omega^2-\w^2_z}\sigma_x, \label{eq:FMarkov} \ee
where $\sigma_z$ and $\sigma_x$ are Pauli matrices. The $F$ matrix is hermitian and traceless, so its two eigenvalues are real and opposite:
\begin{eqnarray}
f_{\pm}(\omega) = \pm \sqrt{1 +\left(\frac{2g^2/\w_z}{\omega} \frac{1}{1- (\omega/\w_z)^2}\right)^2}\;.
\label{eq:fpm}
\end{eqnarray}
Fig. \ref{fig:photon_distribution} shows the behavior of the positive eigenvalue in two points of the phase diagram, nearby and far away from the transition. In both cases, at low frequencies the eigenvalue diverges as $1/\omega$. 
Exploiting the analogy to the equilibrium case, we obtain the LET
\be \widetilde{T}_{\rm eff} = \frac{g^2}{\w_z}.\label{eq:Teff} \ee
We find that $\widetilde{T}_{\rm eff}$ is {\rm not} proportional to the decay rate $\kappa$. The temperature is rather proportional to the effective interaction between the spin and the photon. $g$ is the scale that leads to a competition of unitary and dissipative dynamics, and the LET is a measure of this: For $g=0$, the steady state of the dissipative part of the dynamics (empty cavity) is an eigenstate of the Hamiltonian, while this is no longer the case for any finite $g$. In the limiting case of $g \to 0$, the LET goes to zero.

A closer inspection of (\ref{eq:fpm}) reveals an important difference between the Markovian bath and thermal equilibrium, related to the presence of a second energy scale, $\w_z$.
If $\w_z \gg \widetilde{T}_{\rm eff}$, this energy scale does not affect the  crossover between the quantum and classical regimes, which then proceeds monotonously similar to an equilibrium problem. 
If, on the other hand, $\w_z \ll \widetilde{T}_{\rm eff}$, the quantum-classical crossover of the Markovian bath occurs in an unusual way, highlighting the non-equilibrium nature of the problem. Starting from a divergence at zero frequency, the distribution function (\ref{eq:fpm}) decreases as $\widetilde{T}_{\rm eff}/\w$, in analogy to an equilibrium system at finite temperature. Then, instead of monotonously decreasing towards the quantum regime where $f \approx 1$, it exhibits a second divergence at $\w=\w_z$. 
Since the spectral weight vanishes sufficiently fast in this regime, the correlation functions still remain finite; the pole in $F$ accounts for a different scaling of correlations and spectral properties in this regime (cf. Fig.~\ref{fig:photon_response}). At higher frequencies, it finally tends to one, following the non-equilibrium curve $f \approx \widetilde{T}_{\rm eff}/(\w-\w_z)$. The approach to the quantum regime $f \approx 1$ is polynomial, unlike the exponential approach in the equilibrium case. In App.~\ref{app:1overom}, we show that the thermal $1/\omega$ divergence, leading to a finite LET, is generic for Markovian systems.

We note that our definition of an effective temperature is not the one commonly used in the context of laser theory \cite{scully97}. In this context, the effective temperature is used only to describe the fluctuations of the photonic field, as compared to the equilibrium fluctuations in the lab frame. As a consequence, the divergence of the photon number at the phase transition is always associated to a diverging effective temperature. In contrast, our low-frequency effective temperature (LET) describes the ratio between the fluctuations and the response of the system in the rotating frame. It is finite at the transition and, as we will see, allows us to map the Dicke transition to an existing dynamical universality class of equilibrium systems.


\section{Finite-$N$ corrections from a Keldysh and Langevin perspective}
\label{sec:finite_N}

We now move beyond the quadratic theory, by considering the effects of finite $N$. Based on the formalism developed in the previous section, we approach this problem by scaling analysis, diagrammatic technique, mapping into a low-frequency effective Langevin equation. As will be seen below, these methods show quantitative agreement with a Monte Carlo solution of the original Master equation, highlighting the utility of the present formalism.

%


%
%

\subsection{Scaling analysis}
\label{subsec:finite_size}

Up to this point we have considered only the thermodynamic limit $N\to\infty$. In this limit, the resulting theory is quadratic and can be studied by Keldysh means as well as by the Heisenberg-Langevin method. Corrections due to a finite $N$ introduce non-quadratic terms into the problem and require a more careful study. The present path-integral approach allows us to develop a diagrammatic approach and to resum all leading-order corrections in an organized fashion. A similar approach has been used to study the instability of an optical parametric oscillator in Ref.s \cite{mertens93A,mertens93,feischhauer97}, where however the emerging low-frequency thermal nature of the problem has not been discussed.

\begin{figure}[t]
\includegraphics[scale=0.7]{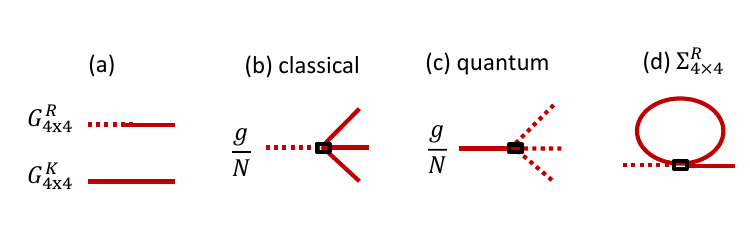}
\caption{ (Color online) Non-equilibrium diagrammatic expansion of the Dicke model to leading order in $1/N$. The dashed lines indicate ``quantum'' fields and the dotted lines ``classical'' fields. (a) Bare Green's function of the photons (red) and of the atoms (blue). (b-c) Leading order $1/N$ corrections: classical vertices contain only one quantum field, while quantum vertices contain more than one (three in this case). (d) One loop correction to the retarded Green's function.}\label{fig:finiteN}
\end{figure}

Leading $1/N$ corrections to the Hamiltonian of the Dicke model are easily obtained by retaining the first-order terms in the Holstein-Primakoff approximation (see Eq.~(\ref{eq:large_N_hamil})). These terms are expressed using the Keldysh formalism in Eq.~(\ref{eq:Squartic}) and contain products of four fields. In a diagrammatic description (see Fig.~\ref{fig:finiteN}), they correspond to four-point vertices. These vertices can be ``classical'' if they contain only one quantum field (either $b_q$ or $a_q$), or ``quantum'' if they contain three of them. The former type can be casted into a semi-classical description of the problem, while the latter describe genuine quantum corrections.

Before going into the calculations, let us first study the relevance (in the sense of the renormalization group (RG)) of the classical and quantum vertices with respect to the critical point in the thermodynamic limit ($N\to\infty$ and $g=g_c$). From the low-frequency expansion of Eq.~(\ref{eq:Sxx}) we obtain that, at this point, the photonic Keldysh action corresponds to:
\be S_{xx} = \int dt~ \left(x_{cl}(t)~x_{q}(t)\right)\left(\ba{c c}0 & -2i\kappa\partial_t \\2i\kappa\partial_t &  8 i \kappa T_{x} \ea\right)\left(\ba{c}x_{cl}(t)\\x_q(t)\ea\right),\label{eq:Sxxlow}\ee
where $T_x$ is defined in Eq.~(\ref{eq:Teff2}). This action is invariant under the scaling transformation 
\bea
t &\to& \lambda ~t \nn,\\
x_{cl}(t) &\to& {\sqrt{\lambda}}~x_{cl}(t)\nn ,\\
x_q(t) &\to& \frac1{\sqrt{\lambda}}~x_q(t).\label{eq:scaling}
\eea
Repeating the same analysis for the atomic field $b$ we again find that, under the scaling transformation, the classical component $b_{cl}$ is increased by a factor $\sqrt{\lambda}$ and the quantum component $b_q$ decreased by the same factor. Using the scaling relation (\ref{eq:scaling}), we find that
\bea 
\frac{g}{N}\int dt~\phi_{cl}\phi_{cl}\phi_{cl}\phi_q &\to& \lambda^2 \frac{g}{N}\int dt~\phi^3_{cl}\phi_q,\label{eq:classicvertex} \\
\frac{g}{N}\int dt~\phi_{cl}\phi_{q}\phi_{q}\phi_q &\to& \frac{g}{N}\int dt~\phi^3_{q}\phi_{cl} .
\eea
Here $\phi$ are one of the $a,~a^*,~b,~b^*$ fields. Eq.~(\ref{eq:classicvertex}) indicates that the classical vertex is relevant in the RG sense, while the quantum vertex is at most marginal. In the limit of $N\gg1$ its contribution is very small at low frequencies and can be neglected. In contrast, the effects of the classical vertex grow as we approach the transition and need to be taken into account.

The above scaling transformation can be used to derive the finite-size scaling of expectation values. For this task, it is convenient to combine the scaling transformation with a renormalization of the system size $N$, such that overall the relevant vertex remains unchanged. Using Eq.~(\ref{eq:classicvertex}) we find that the appropriate transformation is:
\be N \to N' = \frac{N}{\lambda^2} \ee
With this modification, the theory including leading $1/N$ corrections becomes scale invariant at the critical point. Consider now for example the photonic fluctuations $\av{x_{cl}^2}$. This object can be made scale invariant if multiplied by $1/N^{1/2}$, indicating that
\be \av{n} \approx \half\av{x_{cl}^2} \sim N^{1/2}\,. \ee
Remarkably, the same scaling relation holds for the optical parametric oscillator \cite{feischhauer97}, but is here obtained in the framework of the Dicke model. 
In fact, recent numerical calculations on this model \cite{konya12} suggested a different scaling relation $\av{n}\sim N^\alpha$, with $\alpha=0.41$, in contrast to the present analysis.

To further supplement our analytical result, we now consider the effects of next-to-leading-order corrections, stemming from higher-order terms of the Holstein-Primakoff expansion. Their general form is $g/N^k\int dt~\phi^{2+2k}$. For any $k$, the most relevant term is the classical vertex $g/N^k \int dt~ \phi_q\phi_{cl}^{1+2k}$. Under the scaling transformation, this term is multiplied by $\lambda^{1-k}$. This shows that all terms with $k>1$ are irrelevant at a tree level and cannot modify the above scaling relations. 

Before proceeding, we briefly compare the present analysis with the zero temperature equilibrium case. There, the Keldysh component of the action would correspond to $4i\kappa|\omega|$, leading to the scaling transformation $x_{cl} \to \sqrt{\lambda}x_{cl}$ and $x_q \to \sqrt{\lambda}x_q$. As a consequence, both classical and quantum vertices scale in the same manner, and the latter cannot be disregarded. To compute the finite size scaling of expectation values we observe that
\be \frac{g}{N}\int dt~\phi\phi\phi\phi \to \lambda^3 \frac{g}{N'}\int dt~\phi\phi\phi\phi\;. \ee
In order to preserve the scale invariance, we therefore need to renormalize $N$ by
\be N \to N' = \frac{N}\lambda^3 \quad\Rightarrow\quad  \av{n} \sim N^{1/3}\;.\ee
This relation is known in the literature and has been shown to be valid for the zero temperature case, both analytically \cite{vidal06} and numerically \cite{liu09}. As we explained, it does not hold for the (non-equilibrium) thermal case presented here.

\subsection{Diagrammatic calculations}

We now use the Keldysh approach to explicitly compute the photon occupation across the transition in the presence of $1/N$ corrections. As discussed in the previous section, the (bare) Keldysh and retarded propagators of the system are $4\times4$ matrices. In this language, quartic corrections correspond to forth-order tensors of total size $4^4=256$, which we will denote as $\bar{M}$. In our case the relevant corrections are (see Eq. (\ref{eq:Squartic}))
\bea
\frac{g}{4N}&\int_\w& \left[(a_{q} + a_{q}^*)(b_{cl} + b_{cl}^*) + (b_{q} + b_{q}^*)(a_{cl} + a_{cl}^*) \right]  b_{cl} b_{cl}^*\nn\\
&&+\left(b_{q} b_{cl}^*  + b_{q}^* b_{cl}\right)(a_{cl} + a_{cl}^*)(b_{cl} + b_{cl}^*)\;,
\eea
and the tensor $\bar{M}$ contains $16$ identical entries, $\bar{M}_{i,j,k,l}=g/(4N)$.

\begin{figure}[t]
\centering
\includegraphics[scale=0.65]{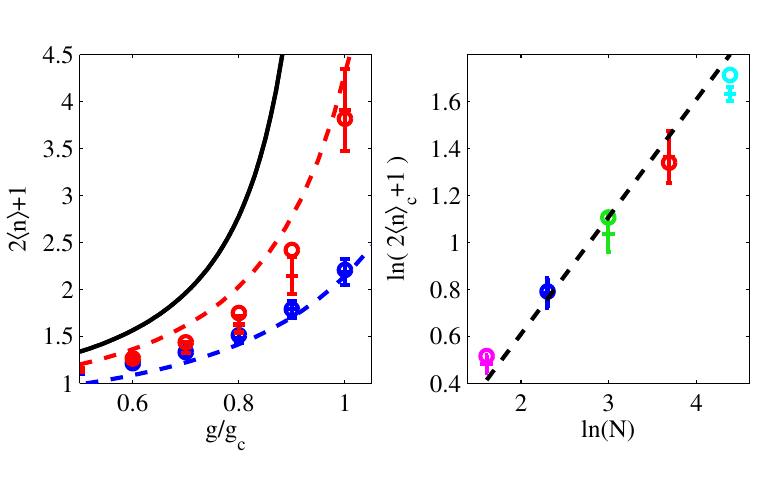}
\caption{ (Color online) {\bf LEFT:} Photon occupation in the vicinity of the Dicke transition. The circles (o) correspond to the one-loop resummation, obtained using the Keldysh diagrammatic technique for $N=10~{\rm (blue)},~40~{\rm (red)}$. The dashed lines correspond to the effective equilibrium theory, Eq. (\ref{eq:finiteNlang}), derived from the Langevin equation.  The crosses (+) correspond to the Monte Carlo solution of the original master equation.  The solid curve corresponds to the mean field solution, Eq.~(\ref{eq:infiniteN}), valid in thermodynamic limit $N\to\infty$. {\bf RIGHT:} Photon occupation at the critical point as function of the system size. The circles (o) and crosses (+) represent respectively diagrammatic and Monte Carlo results. The dashed line corresponds to the effective equilibrium theory (\ref{eq:finiteNlang2}). Numerical parameters: $\w_z=2.0$, $\w_0=1.0$, $\kappa=1.0$, giving $g_c=1.0$.}\label{fig:numerics}
\end{figure}

The leading-order $1/N$ corrections can be computed using standard diagrammatic techniques. When constructing one-loop diagrams one needs to remember that a vertex connects fields at equal time. Because any field $\phi$ satisfies $\av{\phi_q(t)\phi_q(t)}=0$  and $G^R(0) = \av{\phi_q(t)\phi_{cl}(t)}=0$, to obtain a non-vanishing loop one needs to connect two classical edges of the vertex. Thus, one loop corrections renormalize only the retarded and advance Green's function, as shown in Fig.~\ref{fig:finiteN} (d). The analytic expression of the self energy is:
\def \G{{\mathcal G}}
\be 
\Sigma^{R}(\w) = {\mathrm i}\bar{M}\cdot \int d\w' \G^K(\w')  \;,\label{eq:1loop}
\ee
where the operator ``$\cdot$'' indicates the tensorial product including all allowed permutations of the indices. The dressed Green's functions should be computed in a self-consistent manner. The resummation over all one-loop irreducible diagrams leads to the Dyson equation:
\bea
\G^{K}(\w) &=& -\G^R(\w)D^K(\w)\left[\G^R(\w)\right]\yd\\
{[\G^R(\w)]}^{-1} &=& G^{R}(\w) \left[1 + \Sigma^{R}(\w) G^R(\w)\right]^{-1}
\eea
Here $G^{R}$ and $D^K$ are explicitly given in Eq.~(\ref{eq:G4x4}). The resulting predictions for the photon occupation are shown by circles in Fig.~\ref{fig:numerics}.

\subsection{Effective low-frequency Langevin approach and mapping to a thermal ensemble}
\label{subsec:langevin_keldysh}

In this subsection we derive a simple description of the photon-only action (\ref{eq:Sphotononly}), focusing on the $x=(a+a^*)/\sqrt{2\w_0}$ quadrature, by mapping its Keldysh action to a stochastic equation. Using the basic theorems of thermodynamics, we will then convert it into an effective equilibrium free energy, and obtain an analytic expression for the number of photons at the critical point.

We first consider the $N\to\infty$ limit where (as we already saw in Sec.~\ref{sec:prelim} using the Heisenberg-Langevin approach), the Langevin equation coincides with the equation of motion of a classical particle in an harmonic confinement coupled to an equilibrium bath at finite temperature. Using the Keldysh formalism and starting from  Eq.~(\ref{eq:Sphotononly}), we replace $a=\sqrt{\w_0/2}(x+ i p)$ and $a^*=\sqrt{\w_0/2}(x- i p)$  and integrate out the $p$-component, to obtain:
\be S_{xx} = \int_\w \left(x_{cl}(-\w)~x_{q}(-\w)\right)\left(\ba{c c}0 & {[G^A_{xx}(\w)]}^{-1} \\ {[G^R_{xx}(\w)]}^{-1}& D^K_{xx}(\w)\ea\right)\left(\ba{c}x_{cl}(\w)\\x_q(\w)\ea\right),\label{eq:Sxx}\ee
where
\be
D^K_{xx}(\w) = 2 i \kappa\frac{\w_0^2+\kappa^2+\omega^2}{\w_0},\quad\quad {[G^R_{xx}(\w)}]^{-1}= \frac1{\rm{det}{[G^{R}_{2\times2}(\w)]}}, \label{eq:GRKxx}
\ee
and the determinant of $G^{R}_{2\times2}(\w)$ is given by Eq.~(\ref{eq:eigen}). 

As well known, any quadratic Keldysh action is equivalent to a linear Langevin equation. Starting from a generic quadratic action (\ref{eq:Sxx}) one introduces a Hubbard-Stratonovich ``noise'' field $f(\w)$ to obtain:
\begin{align} S_{xf} =\int_\w &2 \Big\{[G^R_{xx}(\w)]^{-1}x_{cl}(\w)-f(\w)\Big\}x_{q}(-\w) - \frac{f(-\w)f(\w)}{D^K_{xx}(\w)}. \label{eq:SHS}\end{align}
(This action is equivalent to Eq. (\ref{eq:Sxx}), as can be explicitly shown by performing the Gaussian integral over $f(\w)$, and using $G^R(\w)=G^A(-\w)$ -- see Ref.~\cite{altlandbook,kamenev_book} for more details.) Inside the Keldysh partition function $Z^K = \int D \{x_q;x_{cl};f\}~e^{i S_{xf}}$, the integration over $x_q$ then takes the form of a delta function with argument
\be [G^R_{xx}(\w)]^{-1} x_{cl}(\w) = f(\w)\label{eq:Langevin_gen}.\ee
The remaining last part of the action Eq. (\ref{eq:SHS}) involves only $f(\w)$ and can be thought of as the statistical weight of a Gaussian random variable with correlations 
\be \av{f(\w) f(\w')} = - i D_{xx}^K(\w)\delta(\w+\w')\,.\label{eq:Langevin_gen2}\ee
Eq. (\ref{eq:Langevin_gen}) then becomes a stochastic equation of motion for the bosonic field $x(\w)=x_{cl}(\w)/\sqrt{2}$, identical to the Langevin equation (\ref{eq:langevin_misha}) obtained in Sec.~\ref{sec:prelim}.

We now include non-linearities for the photon dynamics arising from a finite number of atoms $N$. Our starting point is the frequency limit of Eq.~(\ref{eq:Langevin_gen}), $(2i\kappa\omega + \alpha^2)x(\w)=f(\w)$, where $\alpha$ is defined in Eq.~(\ref{eq:alpha}). To this equation we add the most relevant non-linear term in the from of a frequency independent cubic term: 
\be \Big(2\kappa\partial_t + \a^2\Big) x(t) + \beta^3 x^3(t) = f(t)\label{eq:Langevin2} .\ee
This equation defines the dynamical critical theory of an Ising transition with no conserved quantities, the so-called ``Model A'', for $n=1$ degrees of freedom and $d=0$ dimensions \cite{hohenberg77}. 
The frequency $\beta$ can be determined from the microscopic theory, by demanding the effective Langevin description to reproduce
the same saddle-point as the original action Eq.~(\ref{eq:SDicke}). Using $x=(a+a^*)/\sqrt{2\w_0}$, Eq.~(\ref{eq:alpha}), and Eq.~(\ref{eq:psi0}) we obtain:
\be 
\sqrt{\frac{-\a^2}{\b^3}} = 2\sqrt{N}\frac{\sqrt{2\w_0(g^2-g_c^2)}}{\kappa^2 +\omega^2_0} ~~\Rightarrow~~ \b^3 = \frac{2(\kappa^2+\w_0^2)^2}{N\w_z}.\label{eq:beta}
\ee
As expected, the parameter $\beta$ vanishes in the thermodynamic limit $N\to\infty$, where the mean-field linearized description (\ref{eq:Langevin_gen}) becomes exact.

The stationary state dictated by  the Langevin equation (\ref{eq:Langevin2}) is equivalent to an equilibrium system with free energy
\be F(x) = \half\a^2 x^2 + \frac14\beta^3 x^4\label{eq:Fthermal}\;.\ee
Here we recall that $\alpha$ is defined in Eq.~(\ref{eq:alpha}) and vanishes at the transition, while $\beta$ is defined in Eq.~(\ref{eq:beta}) and captures the $1/N$ corrections. Steady-state expectation values are computed through the thermal average $\langle x^2\rangle = \frac{\int dx~x^2 e^{-F(x)/\widetilde{T}_{\rm eff}}}{\int dx~e^{-F(x)/\widetilde{T}_{\rm eff}}}$. 

In particular, the photon number is:
\bea
2\av{n}+1 &=& 2\w_0\left(1+\frac{\kappa^2}{\w_0^2}\right) \av{x^2}\nn \\
&=& 2\w_0\left(1+\frac{\kappa^2}{\w_0^2}\right) \frac{\int dx~x^2 e^{-F(x)/\widetilde{T}_{\rm eff}}}{\int dx~e^{-F(x)/\widetilde{T}_{\rm eff}}}\label{eq:finiteNlang}\;,
\eea
where in the first identity we used the mean field relation $\av{p^2}=\av{x^2}\kappa^2/\w_0^2$.
At the critical point $\alpha=0$ and the integral is easily evaluated:
\bea 2\av{n}_c+1 &=& 2 \w_0 \left(1+\frac{\kappa^2}{\w_0^2}\right)\sqrt{\frac{\widetilde{T}_{\rm eff}}{\beta^3}} \frac{\Gamma(3/4)}{\Gamma(1/4)}
\nn\\&=&\sqrt{N}\sqrt{\frac{(\kappa^2+\w_0^2)\w_z}{\w_0^3}}\frac{\Gamma(3/4)}{\Gamma(1/4)}.
\label{eq:finiteNlang2} \eea
Here $\Gamma(3/4)/\Gamma(1/4) \approx 0.338$ is the ratio of two Gamma functions.

To evaluate the precision of the Keldysh and Langevin methods, we compare their predictions with the solution of the Master equation associated with the Dicke model (\ref{eq:HDicke1}) with cavity loss (\ref{eq:cav_liou}). Specifically, we apply the Monte Carlo Wave-Function (MCWF) method \cite{molmer93},  as implemented in the open-source C++QED library \cite{C++QED}. (Specific parameters: number of trajectories $N_{\rm traj}=10$, time step $dt=1$, number of time steps $T=400$). The resulting curves are shown in Fig~\ref{fig:finiteN}. We emphasize that no fitting parameters were used when comparing the different methods. As expected, the numerically solution is closer to the predictions of the Keldysh non-equilibrium diagrammatic technique than to the low-frequency thermal effective theory. Remarkably, the difference between these two approaches is minimal at the transition, in agreement with our identification of the transition as driven by equilibrium thermal fluctuations.

\section{Keldysh approach for atom observables}
\label{sec:spontem}

We now analyze the single-atom observables of the open Dicke model using a method which is valid for arbitrary
values of the number of atoms, $N$. To this end, 
we represent each of the $N$ atoms by a real field variable $\phi_\ell$, with the index $\ell$ ranging over all the atoms $\ell=1,..,N$. 
Our method relies on generalizing each $\phi_\ell$ to have $M$ components, $\phi_{a \ell}$ with $a = 1 \ldots M$, and then 
taking the large $M$ limit; even though we are interested in the $M=1$ case, the large $M$ limit is expected to properly describe
the physics of models with long-range interactions \cite{strack11,ye93}. Although we will also consider the large $N$ limit in the present
section in the interest of comparing with previous results, it is important to note that the present method does not require the large $N$ limit,
and is valid for general values of $N$. Also, in the interests of simplicity, we will not write out the $a$ index, and directly
present the large $M$ approximation in the context of the physical $M=1$ case.

This single-atom representation of the Ising spins allows to treat the qualitative effects of atom dissipative dephasing within a simplified ``friction model'' 
for $\phi_\ell$ which we explain below. This process couples directly to the local Ising degrees of freedom of the single atoms. 
 The same is true for disorder due to spatial variations of the qubit-photon couplings \cite{strack11}. 
In such cases, one cannot employ the single large-$N$ Holstein-Primakoff representation of the Dicke model.

We proceed by introducing into the path integral $N$ Lagrange  multipliers $\lambda_\ell$, corresponding to a suitable Fourier representation of the delta function \cite{sachdev_book,strack11}, $\delta(\phi^2_\ell-1)=\int d\lambda \exp^{i\lambda(\phi_\ell^2-1)}$. On the closed time contour, this amounts to adding the following expression to the action:
\begin{align}
S_{\lambda,\pm}&=\frac{-1}{2\w_z} \int_t
\sum_{\ell=1}^N \Bigg[ \lambda_{\ell,+}(t)\left(\phi^2_{\ell,+}(t)-1\right)
-\lambda_{\ell,-}(t)\left(\phi^2_{\ell,-}(t)-1\right)\Bigg].\;
\label{eq:lambda_construction}
\end{align}
Moving to the ``classical/quantum'' notation and adding the bare action of the atoms we obtain:
%
\bea
S_{\phi\phi,\lambda}&=&\frac{1}{\w_z}
\int_{\omega} \sum_{\ell=1}^N
\left(\phi_{\text{cl},\ell} (-\omega)\;
\phi_{\text{q},\ell} (-\omega) \right)
G^{-1}_{\phi\phi,\lambda}
\left( \begin{array}{c}
\phi_{\text{cl},\ell}(\omega)\\ 
\phi_{\text{q},\ell}(\omega)
\end{array} \right)
\nn\\&&+\frac{1}{\w_z}
\int_t \sum_{\ell=1}^N \lambda_{\text{q},\ell}(t),
\label{eq:atom_prop}
\\
G^{-1}_{\phi\phi,\lambda} &=&
\left( \begin{array}{cc}
-\lambda_{\text{q},\ell} & \omega^2 -\lambda_{\text{cl},\ell} + \Sigma^{\text{A}}_{\phi,\ell}(\omega)  \\
\omega^2 -\lambda_{\text{cl},\ell} + \Sigma^{\text{R}}_{\phi,\ell}(\omega) & 
-\lambda_{\text{q},\ell}+\Sigma^{\text{K}}_{\phi,\ell}(\omega)
\end{array} \right)\;.
\nn
\eea
%
Note that $\sqrt{\lambda_{\text{cl},\ell}}$ can be associated with the excitation energy 
of the Ising spins and is to be determined self-consistently. The atom self-energies
$\Sigma^{\text{R/A/K}}_{\phi,\ell}(\omega)$ will be explained below.

Finally, we have for the atom-cavity interaction:
\begin{align}
S_{\phi a}=
\int_t \sum_{\ell=1}^N \frac{g}{2}
\Big[
\phi_{+,\ell}(t) \left(a_{+}(t)+ a^*_{+}(t)\right)
- \phi_{-,\ell}(t) \left(a_{-}(t)+ a^*_{-}(t)\right)
\Big]\;.
\label{eq:S_phia}
\end{align}
The Keldysh action for the full Dicke model (\ref{eq:HDicke1}) then becomes 
\begin{align}
S[a, \phi,\lambda] &= S_{a}+S_{\phi\phi,\lambda}+S_{\phi a}\;,
\label{eq:atom_photon_action}
\end{align}
with the various terms given by Eqs.~(\ref{eq:oscRAKaction},\ref{eq:atom_prop}, \ref{eq:S_phia}). 

We model local, single-atom damping in a simple effective way which is consistent with symmetry properties of our real-valued Ising oscillators $\phi$. The atoms are subject to decay into photon modes outside the 
cavity and possible other damping mechanisms like s-wave scattering with other momentum-modes, trap loss or finite-size 
dephasing \cite{mottl12}. As a result some fraction of the atoms leave the two-density mode Hilbert space 
which maps to the Dicke model; others may be spontaneously scattered back in. 
Representing the atoms  by a complex field 
$\Phi^\ast$, $\Phi$, we subsume the above processes into Markovian decay of the atoms with the self-energies
\begin{align}
\Sigma^{\text{A}}_{\Phi^\ast\text{\MARKOV}}= -i\gamma,\quad
\Sigma^{\text{R}}_{\Phi^\ast\text{\MARKOV}}= +i\gamma,\quad
\Sigma^{\text{K}}_{\Phi^\ast\text{\MARKOV}}= 2 i \gamma,
\label{eq:complex_atom_self_markov}
\end{align}
with $\gamma$ an effective single-atom decay rate. Our effective real-valued Ising field in Eq.~(\ref{eq:atom_prop}) may be viewed as the real component of  the 
originally complex boson $\Phi_{\text{q}/\text{cl}}(t)= \sqrt{\frac{1}2}\left(\phi_{\text{q}/\text{cl}}(t)+i \tilde{\phi}_{\text{q}/\text{cl}}(t)\right)$. Integrating out the $\tilde{\phi}$-component,
\begin{align}
\Sigma^{\text{A}}_{\phi\text{\MARKOV}}(\omega)=-i\gamma \omega ,\quad
\Sigma^{\text{R}}_{\phi\text{\MARKOV}}(\omega)=+i\gamma \omega ,\quad
\Sigma^{\text{K}}_{\phi\text{\MARKOV}}(\omega)= i \gamma~\frac{\omega^2 +\gamma^2 + \w_z^2}{2\w_z}\,\,.
\label{eq:atom_self_markov}
\end{align}
Note that this simple model for dissipative dephasing couples to the $\sigma_x$ projection of the atomic states and does not specify 
the states of the $\sigma_z$ projection of the spins. We emphasize, however, that the form of the dissipative self-energies is dictated by the combination of low-frequency expansion and the real-valued nature of the Ising field $\phi$. In particular, a frequency independent term is ruled out for $\Sigma^{\text{A,R}}$. 
%
%
The above results for Markovian baths should be compared with the results for atoms in equilibrium,
\begin{align}
\Sigma^{\text{A}}_{\phi,\text{\EQUILIB}}(\omega)&=-i\epsilon\omega, \nonumber\quad
\Sigma^{\text{R}}_{\phi,\text{\EQUILIB}}(\omega)=+i\epsilon\omega, \nonumber\\
\Sigma^{\text{K}}_{\phi,\text{\EQUILIB}}(\omega)&=2 i \epsilon\omega \coth\left[\frac{\omega}{2T}\right]\,,
\label{eq:atom_self_equilib}
\end{align}
where one also lets $\epsilon\rightarrow 0$ at the end of the calculation.
We here analyze the Dicke model in terms of the atomic degrees of freedom alone. 
One can exactly integrate out the photons from the action~(\ref{eq:atom_photon_action}). 
This is conveniently done by going to a coordinate representation of the photons:
$a_{\text{q}/\text{cl}}(t)=\sqrt{\frac{\omega_0}{2}} \left(x_{\text{q}/\text{cl}}(t)+i p_{\text{q}/\text{cl}}(t)\right)$,
and first performing the integration over $p_{\text{q/cl}}$ and subsequently over $x_{\text{q/cl}}$. 
We obtain, with $S_{\phi\phi,\lambda}$ given by Eq.~(\ref{eq:atom_prop}), the atom-only action, 
\begin{align}
S[\phi,\lambda] &= S_{\phi\phi,\lambda}+S_{\phi\phi,g^2}\;,
\label{eq:atom_only_action}\\
S_{\phi\phi,g^2}
=&
-\frac{1}{2} \int_\omega \sum_{\ell,m = 1}^N \frac{g^2}{N}\;
\times\label{eq:photon_mediated}\\
&\left(\phi_{\text{cl},\ell} (-\omega)\;
\phi_{\text{q},\ell} (-\omega) \right)
\left( \begin{array}{cc}
0 &\sigma^{\text{A}}(\omega) \\
\sigma^{\text{R}}(\omega)& 
\sigma^{\text{K}}(\omega)
\end{array} \right)
\left( \begin{array}{c}
\phi_{\text{cl},m}(\omega)\\ 
\phi_{\text{q},m}(\omega)
\end{array} \right)
\nonumber\;,
\end{align}
%
where the matrix entries are
\begin{align}
\sigma^{\text{R}}(\omega)&=\left[\sigma^{\text{A}}(\omega)\right]^\ast = \frac{-2\omega_0}{\left(\omega + i \kappa \right)^2 -\omega_0^2} \;,\nonumber\\
\sigma^{\text{K}}(\omega)&=\frac{2 i \kappa \left(\omega^2 + \kappa^2  +\omega_0^2\right)}
{\big |\left(\omega-i\kappa \right)^2 -\omega_0^2\big | ^2} \;.
\label{eq:photon_mediated_2}
\end{align}

Our analysis of the above theory will rely on the approximation of substituting the $N$ Lagrange multipliers by a single effective field $\lambda_{\text{q/cl},\ell}\rightarrow \lambda_{q/cl}$. This ``spherical'' approximation for the Lagrange multiplier becomes exact in the limit of a large number of internal spin components $M\rightarrow\infty$. It can be shown that the critical behavior is not qualitatively modified for any finite value of $M$ including the Ising case $M=1$ of the present paper \cite{ye93}.

The above method is valid for arbitrary values of $N$, and we will describe the general $N$ solution below in Section~\ref{sec:generalN}.
However, first we present a method which efficiently treats the $N \rightarrow \infty$ limit.
We decouple Eq.~(\ref{eq:photon_mediated}) with a Hubbard-Stratonovich field $\psi_\ell(\omega)\leftrightarrow \phi_\ell(\omega)$ and integrate out the $\phi$-field. We assume $\psi$ to be time-independent and spatially uniform $\psi_{\ell}(\w)\rightarrow\psi/(2\pi)\delta_{\w,0}$, and the resulting Keldysh partition function 
\begin{align}
Z^K=\int D\psi D\lambda e^{i\frac{N}{2\pi}\mathcal{S}[\psi,\lambda]}\;,
\end{align}
obtains a prefactor of the number of atoms $N$ in the exponent multiplying the action
\begin{widetext}
\begin{align}
\mathcal{S}[\psi,\lambda]= &\frac{ \lambda_{\text{q}} }{\w_z}
+  g^2\left(\sigma^{\text{R}}(0) \psi_{\text{cl}}\psi_{\text{q}} + \frac{1}{2} \sigma^{\text{K}}(0) \psi_{\text{q}}^2\right)
+\frac{i}2 \int_\omega\left( \ln \Bigg[ 
\lambda_{\text{q}} \left(\Sigma^{\text{K}}_{\phi}(\omega)-\lambda_{\text{q}}\right)
+\left(\omega^2-\lambda_{\text{cl}}+\Sigma^{\text{A}}_{\phi}(\omega)\right)
\left(\omega^2-\lambda_{\text{cl}}+\Sigma^{\text{R}}_{\phi}(\omega)\right)
\Bigg]\right)
 \label{eq:zero_dis_saddle}\\
&- \frac{g^4}4 \frac{\w_z}
{\lambda_{\text{q}}
\left(\Sigma_{\phi}^{\text{K}}(0)-\lambda_{\text{q}}\right)
+
\lambda_{\text{cl}}^2}
\Bigg[
\left(\Sigma^{\text{K}}_{\phi}(0)-\lambda_{\text{q}}\right)\left(\sigma^{\text{R}}(0)\right)^2\psi_{\text{q}}^2
+ 2\lambda_{\text{cl}} \sigma^{\text{R}}(0) \psi_{\text{q}} \left(\sigma^{\text{R}}(0)\psi_{\text{cl}}
+\sigma^{\text{K}}(0)\psi_{\text{q}}\right)
-\lambda_{\text{q}}\left(\sigma^{\text{R}}(0)\psi_{\text{cl}}+ \sigma^{\text{K}}(0)\psi_{\text{q}}\right)^2
\Bigg]\;.\nonumber
\end{align}
\end{widetext}
Taking $N\rightarrow \infty$, we now extract the phase diagram, response and correlation functions, and the value of the order parameter using a saddle-point approximation. This can be obtained by requiring the derivatives with respect to $\lambda_q$ and $\psi_q$ to be zero, and then substituting $\lambda_{\text{q}} = 0$, 
$\lambda_{\text{cl}} =\lambda$, 
$\psi_{\text{q}}=0$,
$\psi_{\text{cl}}=\psi$. 
The derivative with respect to $\lambda_{\text{q}}$ constrains --by construction Eq.~(\ref{eq:lambda_construction})-- the
frequency integral of the Keldysh Green's function to be equal to unity:
\begin{align}
\frac{\partial \mathcal{S}}{\partial \lambda_{\text{q}}}=0\quad \Rightarrow \quad
\av{\phi^2} = \int \frac{ d\omega}{2\pi}i G^{\text{K}}_{\phi\phi}(\omega) = 1\;
\label{eq:constraint}
\end{align}
with
\begin{align}
G^{\text{K}}_{\phi\phi}(\omega)
=&
\frac{-\w_z\Sigma^{\text{K}}_{\phi}(\omega)}
{2\left(\omega^2-\lambda+\Sigma^{\text{A}}_{\phi}(\omega)\right)
\left(\omega^2-\lambda+\Sigma^{\text{R}}_{\phi}(\omega)\right)}
\nonumber\\
&
 - 2\pi\mathrm i \delta(\omega)
\frac{g^2\w_z^2}{4\lambda^2}  \left(\sigma^{\text{R}}(0)\right)^2 \psi^2\;.
\label{eq:keldysh_phi}
\end{align}
The saddle-point condition for the order parameter yields
\begin{align}
\frac{\partial \mathcal{S}}{\partial \psi_{\text{q}}}=0\quad\Rightarrow\quad \psi\left[1-\frac{g^2\w_z}2
\frac{\sigma^{R}(0)}{\lambda}\right]=0\;.
\label{eq:saddle_psi}
\end{align}
To determine the position of the Dicke transition, we need to compute the saddle point value of $\lambda_{cl}=\lambda$ in the normal and in the superradiant phases and equate the two values. 
In the normal (N) phase $\lambda$ is determined by (\ref{eq:constraint}) and (\ref{eq:keldysh_phi}) with $\psi=0$,
\be \lambda_N = \frac{\gamma^2+\w_z^2}{3} \label{eq:lambda_N}.\ee
Note that naively taking $\gamma\rightarrow0$ does not reproduce the equilibrium value for $\lambda_N$, cf. also 
Subsec.~\ref{subsec:compa_atoms}.
In the ferromagnetic phase (FM) the order parameter acquires a finite expectation value $\psi\neq0$ and, to fulfill Eq.~(\ref{eq:saddle_psi}), we need to require the argument of the square bracket to be zero:
\begin{align}
\lambda_{\text{FM}}
&=
\frac{g^2 \omega_0\w_z}{\omega_0^2+\kappa^2},
\label{eq:lambda_FM}
\end{align}
where we have used Eq.~(\ref{eq:photon_mediated_2}) for $\sigma^{\text{R}}(0)$. At the phase boundaries both (\ref{eq:lambda_N}) and (\ref{eq:lambda_FM}) must hold, leading to:
\begin{align}
g^{\text{\MARKOV}}_c = \sqrt{\frac{\left(\gamma^2+\w_z^2\right)\left(\kappa^2+\omega_0^2\right)}{3\omega_0\w_z}}\;.
\label{eq:g_markov}
\end{align}
Therefore, the spontaneous emission weakens the effective photon-atom coupling and shifts the Dicke transititon to large values of the coupling. This effect can be understood in terms of the atom's depolarization, leading to a reduction of the effective number of atoms contributing to the super-radiant transition.

Eqs. (\ref{eq:constraint}) and (\ref{eq:saddle_psi}) determine the value of ferromagnetic order parameter $\psi$ as well. 
When approaching the phase transition from above ($g\ge g_c$), $\psi$ vanishes as
\begin{align}
\psi^{\text{\MARKOV}}= \sqrt{\frac{3(g+g_c)}{4g^2}}\sqrt{g-g_c}\;.
\label{eq:psi_markov}
\end{align}
Compared with a closed system at equilibrium, Eq.~(\ref{eq:psi_equilib}), the order parameter for this open Markovian 
system vanishes with enhanced amplitude but with the same mean-field like square-root exponent.

\subsection{Atom spectral response function}
\begin{figure}[t]
\vspace*{1mm}
\includegraphics[scale=1]{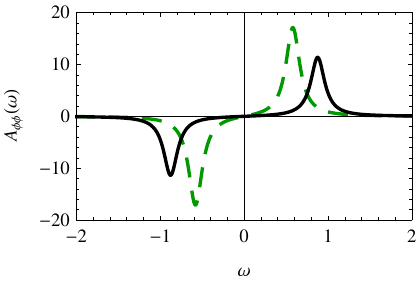}\\[5mm]
\includegraphics*[scale=1]{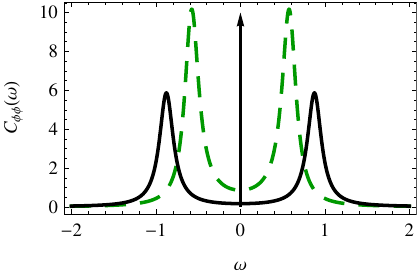}
\caption{(Color online) Atomic spectral response function $\mathcal{A}_{\phi\phi}(\w)$ and 
atomic correlation function $\mathcal{C}_{\phi\phi}(\w)$
in the normal phase ($g<g_c$, green, dashed) and in the superradiant phase ($g=1.2g_c$, black, solid). Other parameters used: $\omega_0 = \w_z = 1$, $\gamma=\kappa=0.2$ (leading to $g_c\approx 0.6$). The arrow illustrates the delta-function contributions from Eq.~(\ref{eq:keldysh_phi}) in the superradiant phase.}
\label{fig:phi_response}
\end{figure}

The single-atom retarded and advanced Green's function are determined by the derivative of (\ref{eq:zero_dis_saddle}) with respect to $\lambda_{\text{cl}}$ and reads
\begin{align}\label{eq:ret_adv_phi}
G^{\text{R}}_{\phi\phi}(\omega)&=\left[G^{A}_{\phi\phi}(\w)\right]^\ast= 
\frac{\w_z}{2\left(\omega^2-\lambda+\Sigma_{\phi}^{\text{R}}(\omega)\right)}\;,
\end{align}
from which follows the spectral response function
\begin{align}
\mathcal{A}_{\phi\phi}(\omega) = -2 \text{Im} G^{\text{R}}_{\phi\phi}(\omega) 
\label{eq:phi_response}
\end{align}
as the expected frequency-resolved signal from the atoms after local, time-modulated density perturbations. Fig.~\ref{fig:phi_response} displays the characteristic Lorentzian shape of the 
spectral response function peaked at $\sqrt{\lambda}$, broadened by single atom decay $\sim \gamma$. 
The single atom response is smooth across the transition. This is not to be confused with the roton-type mode softening 
observed by Mottl {\em et al.} \cite{mottl12} that pertains to the {\em collective} atomic density excitations for a finite number 
of atoms.

\subsection{Atom correlation function}

The atom correlation function is given by the Keldysh Greens function Eq.~(\ref{eq:keldysh_phi})
\begin{align}
\mathcal{C}_{\phi\phi}(\omega)=i G^{\text{K}}_{\phi\phi}(\omega),
\label{eq:phi_spectrum}
\end{align}
$G^K_{\phi\phi}$ is defined in (\ref{eq:keldysh_phi}), and is exhibited in Fig.~\ref{fig:phi_response}. Even before the onset of the superradiance peaks (black and blue-dashed arrows) for $g\geq g_c$, the correlation function 
has finite weight at $\omega=0$. This is the non-equilibrium signature 
of the dissipative dephasing of the pumped atoms due to coupling to the vacuum outside the cavity (a continuum 
of modes with characteristic frequencies orders of magnitudes lower than the optical photons the pumped atoms emit 
when they spontaneously decay and absorb).

\subsection{Comparison with a closed system at equilibrium}
\label{subsec:compa_atoms}

Note that taking $\kappa\rightarrow 0$ and $\gamma\rightarrow 0$ in Eq.~(\ref{eq:g_markov}) does not reproduce 
the equilibrium value for a closed system. This is due to non-commuting limits of making the Markov approximation and 
performing the integral to fulfill the sum rule Eq.~(\ref{eq:constraint}). To obtain $g_c^{\text{\EQUILIB}}$ one needs to use the 
equilibrium bath self-energies 
Eqs.~(\ref{eq:photon_self_equilib},\ref{eq:atom_self_equilib})
from the start of the calculation and obtains the equilibrium analogs of Eqs.~(\ref{eq:g_markov},\ref{eq:psi_markov}):
\begin{align}
g^{\text{\EQUILIB}}_c & = \frac{1}{2} \sqrt{\omega_z\omega_0},\nonumber\\
\psi^{\text{\EQUILIB}}&=\frac{1}{\sqrt{g}}\sqrt{g-g_c}\;.
\label{eq:psi_equilib}
\end{align}
We note that, both dissipative channels, cavity photon loss and atomic dissipative dephasing, shift the critical value of the 
coupling. The amplitude with which the ferromagnetic order parameter vanishes is also different.

Note that in a model for ``one-way'' spontaneous emission coupling to $\sigma^+$ and $\sigma^-$ starting from a fully polarized atomic state as assumed in Sec.~\ref{sec:prelim}, one would recover the equilibrium limit for $\gamma\rightarrow 0$. 
As explained above, our dissipative dephasing model for spontaneous emission couples to $\sigma^x$ and thereby assumes a mixed state of the atoms (similar to a many-body paramagnet).

\begin{figure}[t]
\vspace*{1mm}
\includegraphics*[scale=1]{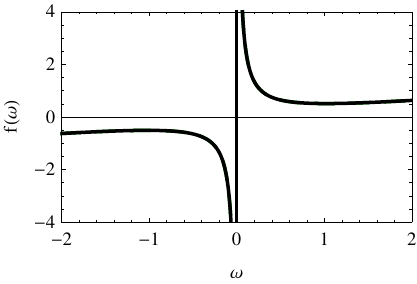}
\caption{(Color online) Distribution function $F_{\phi\phi}(\w)$ for the same numerical parameters as Fig. \ref{fig:phi_response}. The expression for the distribution function Eq.~(\ref{eq:fdrphi}) is independent of the atom-photon coupling $g$.}
\label{fig:phi_distribution}
\end{figure}

\subsection{Atom distribution function and low-frequency effective temperature}
\label{subsubsubsec:eff_temp}

We now execute the procedure of subsection \ref{subsec:eff_temp} to calculate the effective temperature of the 
atoms. With the atom Green's functions and the simple model for atom decay presented above, one finds
\begin{align}
F^{\text{\MARKOV}}_{\phi\phi}(\omega)& = \frac{\mathcal{C}_{\phi\phi}(\omega)}{\mathcal{A}_{\phi\phi}(\omega)}
&=\frac{G^{\text{K}}_{\phi\phi}(\omega)}{G^{\text{R}}_{\phi\phi}(\omega)-G^{\text{A}}_{\phi\phi}(\omega)}
& = \frac{\omega^2 +\gamma^2 + \w_z^2}{2\w_z}\frac{1}{\w},
\label{eq:fdr_phi}
\end{align}
leading to the effective temperature
\begin{align}
T_{\phi}^{\text{eff}}= \frac{\gamma^2+\w_z^2}{4\w_z}\;,
\label{eq:fdrphi}
\end{align}
which is independent of the coupling strength to the photons (cf. also Fig.~\ref{fig:phi_distribution}).
For $\gamma \ll\omega_z$, the effective temperature is set by the recoil energy of the atoms $E_R = \omega_z/2$.
Within our model, $T_\phi^{\text{eff}}$ also does not depend on the cavity loss rate $\kappa$, 
contrary to what obtains for the model of Ref.~\cite{gopa11}. In our case the reason for this is the careful treatment of the thermodynamic limit 
$N\rightarrow\infty$ limit leading to Eqs.~(\ref{eq:atom_only_action},\ref{eq:zero_dis_saddle}). This limit ensures that 
the only photon-induced self-energies for the atoms occur for zero-frequency quantities (the weight of $\delta(\omega)$ 
in Eq.~(\ref{eq:keldysh_phi})).

\subsection{General $N$ solution for the spectral response function}
\label{sec:generalN}

The results presented above refer to the $N\to\infty$ limit. 
However, as we noted earlier, this limit is not really necessary, and the methods of this section can produce general $N$ results
relying {\em only\/} on the $M \rightarrow \infty$ limit. 

We now compute finite size $N$ corrections to the single atom spectral response function, thereby underlining the strength of Keldysh path integrals to perform systematic approximation schemes. It should be noted that, in contrast to the photons' correlations computed in the previous section, the single-atom correlation functions do not diverge at the transition and do not need to be regularized by the number of atoms $N$. Thus, for typical cavity QED experiments where the number of atoms is of the order of $10^5-10^6$, deviations from the $N\rightarrow \infty$ limit will not be observed in the single-atom observables. Nevertheless, the few-body regime might become interesting in future applications.

\begin{figure}[t]
\vspace*{1mm}
\includegraphics*[width=85mm]{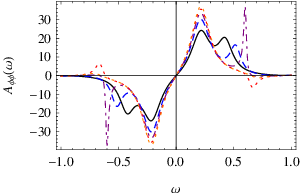}
\caption{(Color online) Finite atom number signatures in the single atom, local spectral response function. Numerical parameters used:
$g=0.4 g_c$, $\kappa=\gamma=0.2$. Line coding: $N=2$, black curve; $N=3$, blue-dashed ; $N=4$ purple-dashed-dotted; $N=5$, red-dashed; $N=6$, orange-dashed. The second peak at $\omega\approx 0.4$ for the black-solid curve is pushed to higher energies until for $N\gtrsim 6$ only the dominant peak at $\omega\approx 0.25$ remains.}
\label{fig:1overN}
\end{figure}

To study the finite-size effects, we write Eq.~(\ref{eq:atom_only_action}) as a RAK matrix of $N\times N$-matrices:
\begin{widetext}
\begin{align}
\int_\omega \sum_{\ell,m=1}^N
\left(\phi_{\text{cl},\ell} (-\omega)\;
\phi_{\text{q},\ell} (-\omega) \right)
\left( \begin{array}{cc}
0 & \frac{\omega^2 -\lambda + \Sigma^{\text{A}}_{\phi}(\omega)}{\omega_z} \delta_{\ell m} -\frac{1}{2} g^2 \sigma^{\text{A}}(\omega) \\
\frac{\omega^2 -\lambda + \Sigma^{\text{R}}_{\phi}(\omega)}{\omega_z} \delta_{\ell m} -\frac{1}{2} g^2 \sigma^{\text{R}}(\omega)  & 
\frac{1}{\omega_z}\Sigma^{\text{K}}_{\phi}(\omega)\delta_{\ell m} - \frac{1}{2} g^2 \sigma^{\text{K}}(\omega)
\end{array} \right)
\left( \begin{array}{c}
\phi_{\text{cl},m}(\omega)\\ 
\phi_{\text{q},m}(\omega)
\end{array} \right)\;,
\end{align}
\end{widetext}
where we have not rescaled the coupling $g$ by $N$ and used the saddle-point values for the other variables. The bottom-left element of the $N\times N$ matrix inverse  gives $G^{\text{R}}_{\phi\phi}(\omega)$, from which follows the spectral response function (see Eq.~(\ref{eq:Aaa}). To invert this matrix we note that all its diagonal and off-diagonal elements are separately equal to each other. Using this property, we obtain that the local response function is:
\begin{align}
G^{\text{R}}_{\phi\phi}(\omega)
=&
\left(1 - \frac{1}{N}\right)\frac{\omega_z}{\omega^2 - \lambda + i \gamma \omega}
+\nn\\
&
\frac{1}{N} \frac{\omega_z}{\omega^2 - \lambda + i \gamma \omega -  \frac{1}{2} N \omega_z g^2 \sigma^{\text{R}}(\omega) }\;.
\end{align}
We observe from this expression that as $N\rightarrow \infty$, only the first term survives, shown in Fig.~\ref{fig:phi_response}.
At finite $N$ an additional mode appears in the single-atom spectrum, which vanishes as $N$ becomes large as shown in 
Fig.~\ref{fig:1overN}. The presence of 
two modes, the atomic and photonic branch, also emerges from an analysis in terms of collective, polaritonic variables \cite{emary03}.

%

\section{Conclusion}
\label{sec:conclusion}
In this paper, we presented a path integral approach for the non-equilibrium steady-states of driven quantum systems coupled to Markovian baths, such as ultracold atoms in optical cavities. In the past, these systems have more often been described using a Master equation formalism. We believe that our Keldysh approach allows an easier comparison with other equilibrium and non-equilibrium (classical and quantum) systems.
While some of the results presented here are actually new, and not just known results re-phrased in a new approach, the full utility of our approach will become clear then computing thermodynamics and critical properties of large, open systems with spatially fluctuating degrees of freedom such as disorder \cite{strack11,muller12}. 
In these correlated quantum many-body situations, Master equation approaches are typically limited to relatively small number of 
atoms and a recipe to compute disorder-averaged quantities does not seem to exist.

We first applied our formalism to the cavity vacuum (Sec.\ref{sec:cavityonly}) and subsequently added atomic qubits, interacting with the cavity through a Dicke interaction, and computed the key observables for both the photons (Sec. \ref{sec:dicke_kappa} and \ref{sec:finite_N}) and atoms (Sec.\ref{sec:spontem}).  The key novelties of our analysis are:\\
(i) The fluctuation-dissipation relation of a single cavity coupled to a Markovian bath in the rotating frame, Eq.~(\ref{eq:F_ratio}), differs from the thermal-equilibrium case. In the former case the bath contains both positive and negative frequency, while in the latter it can contain only positive frequency, leading to a different symmetry with respect to $\w\to-\w$.\\
(ii) Nevertheless, in the presence of a drive, the low-frequency distribution functions of the photons and atoms is thermal-like and diverges as $\sim 1/\omega$, allowing the definition of a low-frequency effective temperature (LET). The LET of the photons, Eq.~(\ref{eq:Teff}), and of the atoms, Eq.~(\ref{eq:fdrphi}), are however different, highlighting the non-equilibrium nature of the problem. \\
(iii) At higher frequencies, the distribution functions display non-equilibrium and quantum behaviors. For example, the photon distribution contains a gapped mode, Eq.~(\ref{eq:thetastar}), whose quantum fluctuations remain identical to the zero temperature case throughout the transition.\\
(iv) The thermal-like divergence of the distribution functions determines the critical properties of the ``superradiant'' phase transition. In particular, the photon number diverges as $1/|g-g_c|$ for $N\to\infty$ and scales as $N^{1/2}$ for $g=g_c$. Both results coincide with the equilibrium behavior of a Landau-Ginzburg model at finite temperature, Eq.~(\ref{eq:Fthermal}), and differ from the well studied zero-temperature case (where one obtains $1/|g-g_c|^{1/2}$ and $N^{1/3}$).\\
(v) Dissipative dephasing processes involving single atoms can also be studied using non-perturbative techniques. As long as the symmetries of the original model are preserved, a Dicke transition is still expected, but its position may be strongly renormalized even for small decay rates, Eq.~(\ref{eq:g_markov}), due to the depolarization of the atomic ensemble.\\
(vi) Within the nonlinear sigma model approach (Section~\ref{sec:spontem}), we can obtain the spectral properties of the single atoms for general finite values of $N$ across the phase transition in the dissipative Dicke model.

In the future, it will be interesting to apply our approach to dissipative quantum glasses coupled to Markovian (and other) baths such as potentially 
achievable in multi-mode optical cavities \cite{gopa11,strack11,muller12} or circuit QED \cite{filipp11}. 
It would also be desirable to obtain a more general classification of conditions under which quantum phase transitions of closed systems are turned into thermal phase transitions by dissipation--and perhaps to find counterexamples by engineered dissipation along the lines of Refs.~\cite{schwuehl08,diehl10}.

\begin{acknowledgments}

We thank M. A. Baranov, F. di Piazza, S. Gopalakrishnan, B. Halperin, E. Kessler, D. Marcos, J. Otterbach, P. Rabl, H. Ritsch, L. Sieberer, H. Tureci, and P. Zoller for useful discussions and M. Buchhold, H. Ritsch, D. Marcos and S. Gopalakrishnan for critical comments on the manuscript.

This research was supported by the U.S. National Science Foundation under grant DMR-1103860, by the U.S. Army Research Office Grant W911NF-12-1-0227, by the Center for Ultracold Atoms (CUA), by the Multidisciplinary University Research Initiative (MURI), by the Packard Foundation, by the DARPA OLE program, by the DFG under grant Str 1176/1-1, by the Austrian Science Fund (FWF) through SFB FOQUS F4016-N16 and the START grant Y 581-N16, by the European Commission (AQUTE, NAMEQUAM), and by the Institut f\"ur Quanteninformation GmbH.

\end{acknowledgments}

\begin{appendix}

\section{Self-energy of an open cavity in the rotating frame}
\label{app:integration}

In this appendix, we discuss how the coherent drive with a frequency scale $\omega_p$, which exceeds all other frequency scales, justifies the form of the Markovian dissipative action Eq.~\eqref{eq:Scavity}, which in particular displays frequency independent terms only which are $\delta$-correlated in time and neglect memory effects.

Our starting point is the Hamiltonian of a single boson $ a$, coupled to a continuum of vacuum fields $\psi_k$, via
\be H_0 = \w_c  a\yd  a+ \sum_k \w_k \psi\yd_k \psi_k + g_k ( a \yd \psi_k +  a \psi\yd_k) .\label{eq:Hcavity}\ee
Here $g_k$ is the coupling constant between the cavity boson and the external vacuum, and we neglected counter-propagating terms of the form $a\yd \psi_k\yd$.
Eq.~(\ref{eq:Hcavity}) is quadratic in $\psi_k$, allowing us to analytically integrate-out the vacuum fields and obtain a cavity-only action of the form (\ref{eq:oscRAKaction}). If we assume that the vacuum fields are kept at an equilibrium temperature $T_{\rm ext}=300K$, the corresponding entries are
\bea
[G^R(\w)]^{-1} &=& \w-\w_c -\delta\w + i K(\w),\nn\\
D^K &=& 2 i K(\w) \coth \left(\frac{\w}{2 T_{\rm ext}}\right). \label{eq:GRKeq}
\eea
Here $\delta\w$ corresponds to the Lamb shift and can be absorbed in a finite renormalization of $\w_c$. The function $K(\w) = \sum_k |g_k^2|\delta(\w-\w_k)$ is the spectral density of the vacuum. 

The inverse Green's functions (\ref{eq:GRKeq}) describe the cavity mode in the lab frame. In practice, it is often more convenient to move to a frame rotating with a constant frequency, in our case corresponding to the pump frequency $\w_p$. In this frame, the photons are described by Eq. ($\ref{eq:GRKeq}$) with $\w \to \w_p+\w$. Next, we apply the equivalent of the Wigner-Weisskopf approximation, by Taylor expanding the inverse Green's function in small $\omega$ to zero order. This approximation is justified by the energy scale separation discussed in the text. We obtain
\bea
[G^R(\w)]^{-1} &=& \w-\w_c -\delta\w-\w_p + i K(\w_p),\nn\\
D^K &=&2i  K(\w_p) \coth \left(\frac{\w_p}{2 T_{\rm ext}}\right) .\label{eq:GRKneq}
\eea
The factor $\coth \left(\frac{\w_p}{2 T_{\rm ext}}\right)$ plays the role of the $2n+1$ factor appearing in the finite temperature extension of the master equation (\ref{eq:master-gen}). To be precise, the two expressions coincide only for $\w_c=\w_p$. 
For most experiments the pump frequency is anyway much higher than the external temperature and we can approximate $\coth \left(\frac{\w_p}{2 T_{\rm ext}}\right)=1$. Under this approximation, Eq. (\ref{eq:GRKneq}) becomes equivalent to Eqs.~(\ref{eq:oscGRAK}-\ref{eq:photon_self_markov}) with
\bea 
\w_0 &=& \w_c+\delta\w-\w_p\;,\hspace{5mm}
\kappa = K(\w_p).
\eea

\section{Photon-only action in the superradiant phase}
\label{app:superradiant}

In the superradiant (SR) phase, the field $a$ ($b$) is given by the sum of a time-independent component $a_0$ ($b_0$) and a fluctuating term. The action governing the fluctuating terms can be obtained from Eq.~(\ref{eq:SDicke}) by substituting $a \to a_0 + \delta a$ ($b \to b_0 + \delta b$) (we choose $a_0,b_0$ real without loss of generality). At $N\rightarrow \infty$, we end up with the quadratic action
\begin{eqnarray}\label{eq:Ssuper}
S_{\rm SR} =\frac{1}{2}\int_{\omega} \delta V_{8}^\dag(\omega)
\left(\begin{array}{cc}
0 & [G_{4\times4}^A]^{-1}(\omega) \\
 {{[G_{4\times4}^R]}}^{-1}(\omega) &  D_{4\times4}^K
\end{array}\right)\delta V_{8}(\omega)
\nonumber\\
\end{eqnarray}
with the Green's functions ${[G_{4\times4}^R]}^{-1}(\w) =\left({[G_{4\times4}^A]}^{-1}(\w)\right)\yd = $ 
\begin{eqnarray}
 \left(\begin{array}{cccc}
\omega -\w_0 + i\kappa & 0  & -\bar{g} & -\bar{g} \\
0 & -\omega - \omega_0 -i\kappa & -\bar{g} & -\bar{g} \\
-\bar{g} & -\bar{g} & \omega - \omega_z -2\delta\w_z  & -\delta\w_z  \\
-\bar{g} & -\bar{g} & -\delta\w_z& - \omega - \omega_z - 2\delta\w_z  \\
\end{array}\right),\nonumber
\end{eqnarray}
\bea
D^{K}&=& 2 \mathrm i \,\,\mathrm{diag} (\kappa,\kappa,0,0).
\eea
Here the 8-vector $\delta V_{8}(\w)$ is defined in the analogous way to $V_{8}(\w)$ of Eq.~(\ref{eq:V_8}) 
and 
\begin{align} 
\bar{g} &= g - \frac{3g}N b_0^2 \approx \frac{3g_c^2+g^2}{4g}\;, \\
\bar{\delta \w_z} &= \frac{4g}N b_0(a_0+a_0^*) \approx -4\w_z\frac{g^2-g_c^2}{g^2}.
\end{align}

We note that the principal change to the spectral response and correlation function in the superradiant 
phase 
\bea
\mathcal A_{aa\yd}(\w) &=&\mathcal A_{\delta a,\delta a\yd}(\w)\;, \\
\mathcal C_{aa\yd}(\w) &=&\mathcal C_{\delta a,\delta a\yd}(\w) +  a_0^2 \delta_{\w,0}
\eea
is the $\delta$-function peak at $\omega=0$ in the correlation function due to coherent photons (``photon condensate'').
\section{Damped dynamics near the phase transition}
\label{app:natoverdamp}

We argue based on a systematic low-frequency expansion of the inverse retarded Green's function that the overdamped dynamics observed in the vicinity of the phase transition is generic for systems where a phase transition is driven by a competition within the Hamiltonian sector, while dissipative dynamics acts as a ``spectator''. To see this, we (i) write the most general form of the inverse retarded Green's function 
\begin{align}
&{[G^{R}_{2\times2}]}^{-1}(\omega) = \left(\ba{c c} p(\omega) & o(\omega)
\\o^*(-\omega)  & p^*(-\omega) \ea\right)\;,
\label{eq:GRphoton2}
\end{align}
and (ii) use that the phase transition is governed by low frequency behavior and an expansion in powers of the frequency is appropriate, 
\begin{eqnarray}
p(\omega) =  -\nu + z \omega , \quad  o (\omega) = - \mu + y \omega
\end{eqnarray} with complex coefficients and a low frequency spectrum
\begin{eqnarray}\label{eq:massgap}
\omega_{\pm} = \frac{ \mathrm i \mathrm{Im}[z^*\nu - y^*\mu ] \pm \sqrt{(|z|^2 - |y|^2) \alpha^2 - (\mathrm{Im}[z^*\nu - y^*\mu ])^2}}{(|z|^2 - |y|^2) }.\nonumber\\
\end{eqnarray} 
Without dynamic renormalization effects, $z =1$ and all other frequency coefficients are zero, so they will be generically much smaller than one (more precisely, $|\mathrm{Im}[ z]|,|\mathrm{Re} [y]|,|\mathrm{Im} [y]|\ll1$), and in particular $|z|^2 \gg |y|^2$. (In the large $N$ open Dicke model, they are exactly zero.) A mass gap, i.e. the scale that characterizes the action at zero frequency, provides a measure of the distance from the phase transition and reads 
\begin{eqnarray}
\alpha^2  \equiv  \det G^{R\, -1}_{2\times2} (\omega =0) = |\nu|^2 - |\mu|^2 \geq 0
\end{eqnarray} 
(the last inequality must hold for a stable physical system). Approaching the phase transition, this gap shrinks to zero, such that the frequencies must become purely imaginary as a generic feature of a phase transition in the presence of dissipation.  
Indeed, in a situation where the phase transition is driven by a competition within the Hamiltonian sector of the problem by a quantity $g$, the dominant $g$ dependence is contained in $\alpha^2(g)$ (more precisely, $ \mathrm{Re} [\nu],\mathrm{Re}[\mu]$), while the dissipative scales ($ \mathrm{Im} [\nu],\mathrm{Im}[\mu]$) do not strongly depend on $g$ and remain essentially at their bare, finite values even at the transition point. (In the open Dicke model, only the real parts are modified, while the imaginary parts exactly remain at their bare values.) In such a situation, as $\alpha(g \to g_c) \to 0$, we may expand the square root in Eq. \eqref{eq:massgap} in $\alpha^2$, identifying that parameter as the distance from the phase transition. 


\section{$\boldsymbol{1/\omega}$ divergence in Markov distribution functions}
\label{app:1overom}

We here show that the $1/\omega$ pole in the photon distribution function at low frequency, and the associated low-frequency effective temperature (LET), is indeed a generic feature of Markovian non-equilibrium systems. 

To this end, we consider the low-frequency regime, where in the spirit of a systematic derivative expansion the inverse retarded Green's function of the photon takes the form
\begin{eqnarray}
G^{R\,-1} = (\omega + \mathrm i \nu_2) \sigma_z - H , \quad H  = \nu_1 \mathbf 1 + \mu_1 \sigma_x + \mu_2 \sigma_y,
\end{eqnarray}
where $\nu_1,\mu_1 (\nu_2,\mu_2)$ denote the real (imaginary) part of $\nu,\mu$. We set $z =1,y=0$ here, which in principle contribute at $\mathcal O(\omega)$, and anticipate that this omission will not alter the qualitative results. The hermitean part $H$ represents Hamiltonian dynamics, the antihermitean $\sim \mathrm i \sigma_z$ decay. In a derivative expansion, the most general form of the Keldysh component is 
\begin{eqnarray}
D^K  = 2 \mathrm  i (\kappa_1 \mathbf{1} + \kappa_2 \sigma_x) .
\end{eqnarray}
Solving the fluctuation-dissipation relation, we obtain
\begin{eqnarray}
F(\omega) = \frac{\kappa_1}{\nu_2} \sigma_z - \frac{1}{\omega} \Big[ \frac{\kappa_1 }{\nu_2}  \Big( \mu_1 \sigma_x +   \mu_2 \sigma_y \Big) + \frac{\kappa_2 \mu_1 }{\nu_2} \mathbf{1} + \kappa_2 \sigma_y   \Big].\quad
\end{eqnarray}

Crucially, this confirms the $1/\omega$ divergence behavior of the distribution function. Allowing for the most general form of $G^{R\,-1}$ of the frequency expansion in terms of a finite imaginary part of $z$ and finite $y$ only results in subleading corrections: also in this case, $\lim_{\omega \to 0} [\omega \cdot F(\omega)] \to \mathrm{const}.$ Clearly, adding frequency dependent terms to $D^K$ only leads to subleading corrections in $F$. Therefore, the $1/\omega$ pole at low frequency, and the associated scale generated in this regime, the LET, is a generic feature of Markovian non-equilibrium systems.

\end{appendix}

\bibliographystyle{aipnum4-1}
\bibliography{Dicke}

\end{document}